\documentclass[prb, twocolumn, showpacs, amsmath, amssymb,superscriptaddress, floatfix]{revtex4-2}

\usepackage{amssymb,amsfonts,amsmath} 
\usepackage{graphicx}
\usepackage{color}
\usepackage{hyperref}
\usepackage{tikz}
\DeclareMathOperator{\Tr}{Tr} 
\usepackage{comment}
\usepackage{nicefrac}
\usepackage{braket}
\usepackage{sidecap}
\sidecaptionvpos{figure}{l}

\newcommand{\midarrow}{\tikz \draw[ -stealth] (0,0) -- +(.1,0);}
\newcommand{\midarrowrev}{\tikz \draw[stealth -] (0,0) -- +(.1,0);}
\newcommand{\middownarrow}{\tikz \draw[stealth -] (0,0) -- +(0,.1);}
\newcommand{\miduparrow}{\tikz \draw[-stealth ] (0,0) -- +(0,.1);}

\tikzset{->-/.style={decoration={
  markings,
  mark=at position .6 with {\arrow{stealth}}},postaction={decorate}}}

\newcommand{\Pch}{
\tikz {
\draw [fill = gray] (0.5,0) rectangle (0.75,1);
\draw (0,0)  --node {\midarrow}(0.5,0) ;
\draw (0,1) --node {\midarrow}(0.5,1) ;
\draw (0.75,0)  --node {\midarrow}(1.25,0) node[] at (1.25, -0.5) {$P^{\Lambda}$};
\draw[dashed] (0.75,1) --node {\midarrow}(1.25,1) ;

\draw [fill = gray] (1.75,0) rectangle (2.,1);
\draw (1.25,0)  --node {\midarrow}(1.75,0) ;
\draw[dashed] (1.25,1) --node {\midarrow}(1.75,1) ;
\draw (2,0)  --node {\midarrow}(2.5,0);
\draw (2,1) --node {\midarrow}(2.5,1);
}}

\newcommand{\Cch}{
\tikz {
\draw [fill = gray] (0.5,0) rectangle (0.75,1);
\draw (0,0)  --node {\midarrowrev}(0.5,0) ;
\draw (0,1)  --node {\midarrow}(0.5,1) ;
\draw (0.75,0)  --node {\midarrowrev}(1.25,0) node[] at (1.25, -0.5) {$C^{\Lambda}$};
\draw[dashed] (0.75,1) --node {\midarrow}(1.25,1) ;

\draw [fill = gray] (1.75,0) rectangle (2.,1);
\draw (1.25,0)  --node {\midarrowrev}(1.75,0) ;
\draw[dashed] (1.25,1) --node {\midarrow}(1.75,1) ;
\draw (2,0)  --node {\midarrowrev}(2.5,0);
\draw (2,1) --node {\midarrow}(2.5,1);
}}

\newcommand{\Dch}{
\tikz {
\draw [fill = gray] (0.5,0.5) rectangle (0.75,1);
\draw (0,1) --node {\midarrow}(0.5,1) ; 
\draw (0.75,1) --node {\midarrow}(1.25,1) ;
\draw (0.5,0.5) arc (90:180:0.25) node{\middownarrow} arc (180:270:0.25);
\draw[dashed] (0.75,0.5) arc (90:0:0.25) node{\miduparrow} arc (0:-90:0.25);
\draw [fill = gray] (0.5,0) rectangle (0.75,-0.5);
\draw (0.0,-0.5)  --node {\midarrow}(0.5,-0.5) ;
\draw (0.75,-0.5)  --node {\midarrow}(1.25,-0.5);

\node[] at(1.75,0.25){$+$};
\node[] at(4.62,0.25){$+$};

\draw [fill = gray] (3.1,1) rectangle (3.35,0.25);
\draw (2.25,1)  --node {\midarrow}(3.1,1) ; 
\draw (3.35,1) --node {\midarrow}(4.25,1) ;

\draw[->-] (2.75,-0.25) -- (3.1,0.25);
\draw[->-, dashed] (3.35,0.25) -- (3.75,-0.25);

\draw [fill = gray] (2.75,-0.5) rectangle (3.75,-0.25);
\draw (2.25,-0.5)  --node {\midarrow}(2.75,-0.5) ;
\draw (3.75,-0.5)  --node {\midarrow}(4.25,-0.5);

\draw [fill = gray] (5.5,1) rectangle (6.5,0.75);
\draw (5,1)  --node {\midarrow}(5.5,1);
\draw (6.5,1)  --node {\midarrow}(7,1);

\draw[->-] (5.5,0.75) -- (5.85,0.25);
\draw[->-, dashed] (6.1,0.25) -- (6.5,0.75);

\draw [fill = gray] (5.85,-0.5) rectangle (6.1,0.25);
\draw (5,-0.5) --node {\midarrow}(5.85,-0.5) ; 
\draw (6.1,-0.5) --node {\midarrow}(7,-0.5);
\draw [decorate,decoration={brace,amplitude=10pt,raise=4pt},yshift=0pt]
(7,-0.5) -- (0,-0.5) node [black,midway,yshift=-0.8cm] {$D^{\Lambda}$};
}}

\usetikzlibrary{arrows,shapes,positioning}
\usetikzlibrary{decorations.markings}
\usetikzlibrary{decorations.pathmorphing}
\usetikzlibrary{decorations.pathreplacing}

\usetikzlibrary[arrows]
\tikzstyle{arrowstyle}=[scale=1]
\usepackage{tkz-euclide}

\newlength{\overwritelength}
\newlength{\minimumoverwritelength}
\newcommand{\overwrite}[3]{%
  \settowidth{\overwritelength}{$#1$}%
  \ifdim\overwritelength<\minimumoverwritelength%
    \setlength{\overwritelength}{\minimumoverwritelength}\fi%
  \stackrel
    {%
      \begin{minipage}{\overwritelength}%
        \color{#2}\centering\small #3\\%
        \rule{1pt}{9pt}%
      \end{minipage}}
    {\colorbox{#2!30}{\color{black}$\displaystyle#1$}}}

\usetikzlibrary{arrows,shapes,positioning}
\usetikzlibrary{decorations.markings}
\usetikzlibrary{decorations.pathmorphing}
\usetikzlibrary{decorations.pathreplacing}

\usetikzlibrary[arrows]
\usetikzlibrary{fadings,shapes.arrows,shadows}
\tikzstyle{arrowstyle}=[scale=1]

\begin{document}

\title{Electronic instabilities in Penrose quasicrystals:\\
competition, coexistence and collaboration of order}

\author{J.B.\ Hauck}
\affiliation{Institute for Theoretical Solid State Physics, RWTH Aachen University, 52074 Aachen, Germany}
\author{C.\ Honerkamp}
\affiliation{Institute for Theoretical Solid State Physics, RWTH Aachen University, 52074 Aachen, Germany}
\affiliation{JARA-FIT, J{\"u}lich Aachen Research Alliance - Fundamentals of Future Information Technology, Germany}
\author{S.\ Achilles}
\affiliation{J{\"u}lich Supercomputing Centre, Forschungszentrum J{\"u}lich, Wilhelm-Johnen-Straße, 52425 J{\"u}lich, Germany}
\affiliation{RWTH Aachen University, Aachen Institute for Advanced Study in Computational Engineering Science, Schinkelstr.~2, 52062 Aachen, Germany}
\author{D.M.\ Kennes}
\affiliation{Institut f{\"u}r Theorie der Statistischen Physik,  RWTH Aachen,  52074 Aachen,Germany and JARA - Fundamentals of Future Information Technology}
\affiliation{Max Planck Institute for the Structure and Dynamics of Matter and Center for Free Electron Laser Science, 22761 Hamburg, Germany}

\begin{abstract} 
 Quasicrystals lack translational symmetry, but can still exhibit long-ranged order, promoting them to candidates for unconventional physics beyond the paradigm of crystals.
 Here, we apply a real-space functional renormalization group approach to the prototypical quasicrystalline Penrose tiling Hubbard model treating competing electronic instabilities in an unbiased, beyond-mean-field fashion.  Our work reveals a delicate interplay between charge and spin degrees of freedom in quasicrystals. Depending on the range of interactions and hopping amplitudes, we unveil a rich phase diagram including antiferromagnetic orderings, charge density waves and subleading, superconducting pairing tendencies. For certain parameter regimes we find a competition of phases, which is also common in crystals, but additionally encounter phases coexisting in a spatially separated fashion and ordering tendencies which mutually collaborate to enhance their strength. We therefore establish that quasicrystalline structures open up a route towards this rich  ordering behavior uncommon to crystals and that an unbiased, beyond-mean-field approach is essential to describe this physics of quasicrystals correctly.

\end{abstract}

\date{\today} 
\maketitle

\emph{Introduction. --- }
The discovery of quasicrystals has triggered exciting, pioneering experimental~\cite{bursill_penrose_1985, shechtman_metallic_1984,shi_frustration_2018,deguchi_quantum_2012, yao_quasicrystalline_2018, deng_interlayer_2020, collins_imaging_2017} and theoretical~\cite{steurer_quasicrystals_2018, inayoshi_excitonic_2020, fujiwara_localized_1988, jagannathan_magnetic_1997, moon_quasicrystalline_2019,sakai_effect_2020, rai_proximity_2019, rai_superconducting_2020, lifshitz_photonic_2005, lifshitz_magnetically-ordered_2004} research on the topic. Recently, in experimental studies, superconductivity~\cite{kamiya_discovery_2018} as well as antiferromagnetic ordering~\cite{yoshida_antiferromagnetic_2019} were observed in quasicrystalline systems and their approximants, which are large clusters of quasicrystalline tilings as periodically repeated unit cells. These new experimental findings, especially the reported superconductivity, cannot be explained by our current theory and, therefore, new theoretical approaches are needed. 
For theory, an intrinsic complication of quasicrystals is the non-layered structure in three dimensions (3D). An exception to this are twisted materials, like twisted bilayer graphene,  which form a quasiperiodic lattice in the $x-y$ plane projection for a variety of incommensurable twisting angles~\cite{yao_quasicrystalline_2018}. Additionally, the lack of translational symmetry in quasicrystals leads to a loss of momentum conservation resulting in severe computational challenges, necessitating the use of simplifying models. 

A much-studied example, and one of the prototype models of a two-dimensional (2D) quasiperiodic structure is the Penrose model (a Hubbard model on a Penrose tiling)~\cite{penrose_pentaplexity_1979,hubbard_electron_1963}, which we examine in this letter. It has a single point of global five-fold rotational symmetry and a local ten-fold rotational symmetry~\cite{lifshitz_symmetry_2011}. The Penrose tiling can be seen as the 2D cut through an icosahedral quasicrystal~\cite{bursill_penrose_1985} and constitutes a prototypical model to understand many phenomena in quasicrystalline materials on a qualitative basis, e.g., being currently used to investigate the experimentally observed superconductivity~\cite{sakai_superconductivity_2017,araujo_conventional_2019,cao_kohn-luttinger_2020,sakai_exotic_2019,chen_topological_2019,takemori_physical_2020}, although a direct connection to the three dimensional materials is less clear. Other platforms which can be used to more faithfully realize the $2$D Penrose model are quantum simulators~\cite{mace_quantum_2016} using ultra-cold atomic gases. Possible explanations for the experimentally reported superconductivity in quasicrystals include the existence of unconventional superconducting order generated by spin fluctuations~\cite{cao_kohn-luttinger_2020}. However, the mean-field (MF) theory employed in these studies is biased due to the choice of the decoupling and it remains unclear whether other orderings might prevail.
Thus, there is clear demand for an unbiased, beyond-MF study to explore the physics of quasicrystals.

In this letter, we systematically study the electronic instabilities of the quasicrystalline Penrose model employing such a beyond-MF method, the here developed real-space truncated unity FRG (TUFRG)~\cite{lichtenstein_high-performance_2017,weidinger_functional_2017}. The real-space TUFRG offers a versatile tool for the study of translation symmetry-broken models, such as quasicrystals, and scales favorably enough to reach the thermodynamic limit. Utilizing this advance we unveil the surprisingly rich ordering behavior of quasicrystals expanding significantly on what is realized in crystals. Studying different Penrose models and parameters, we find either a mutual suppression of order, similar to a competition of phases known from crystals, or mutual evasion of order as well as mutual collaboration of multiple ordering tendencies. The latter two are both usually not found in crystals. With this we expand the catalog of how phases of matter emerge in quasicrystals  and show that in general one requires an unbiased, beyond-MF approach to capture the intricate nature and interplay of orderings in these systems.

\begin{figure}
       \includegraphics[width = 0.49\columnwidth]{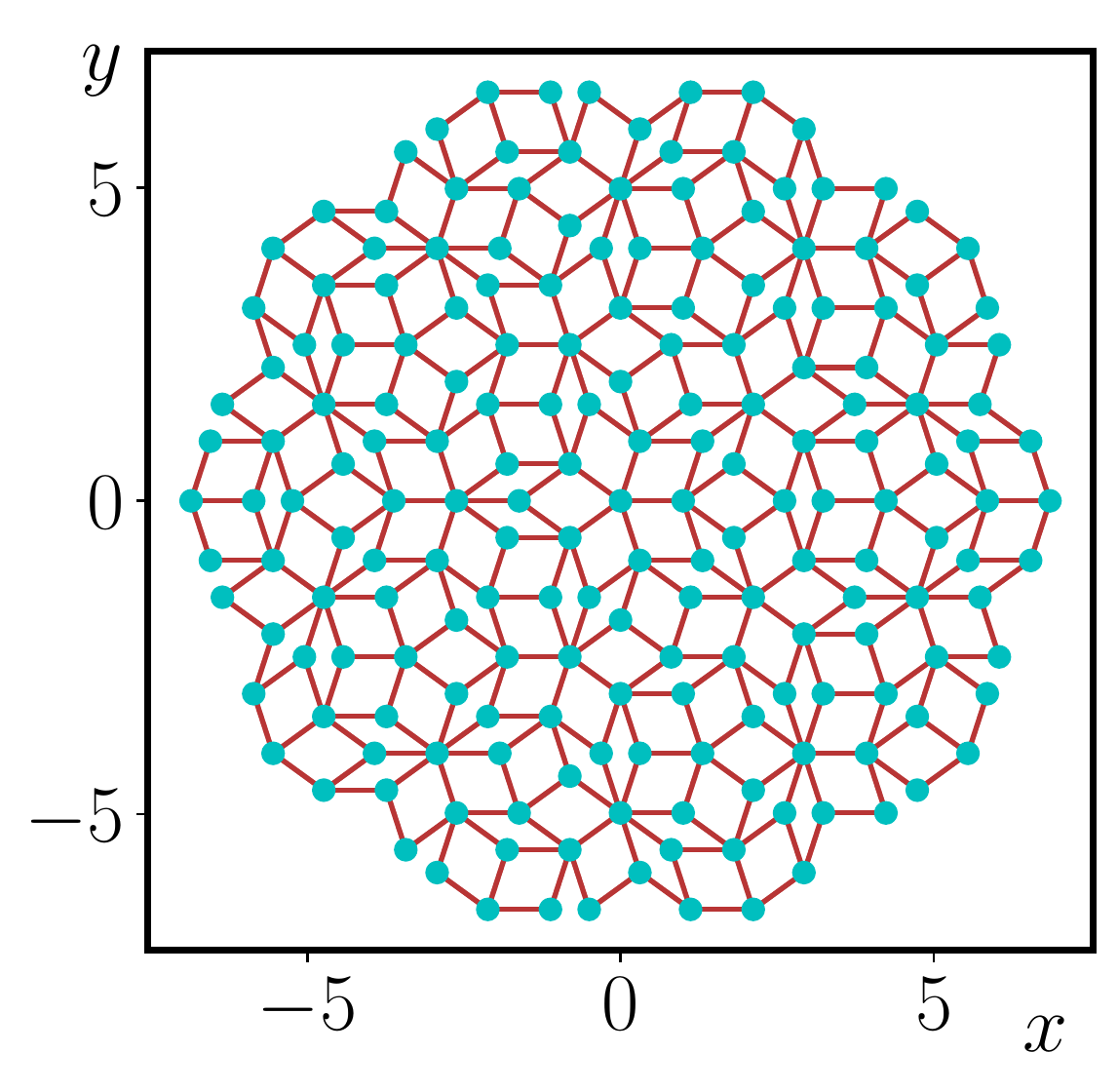}
        \includegraphics[width = 0.49\columnwidth]{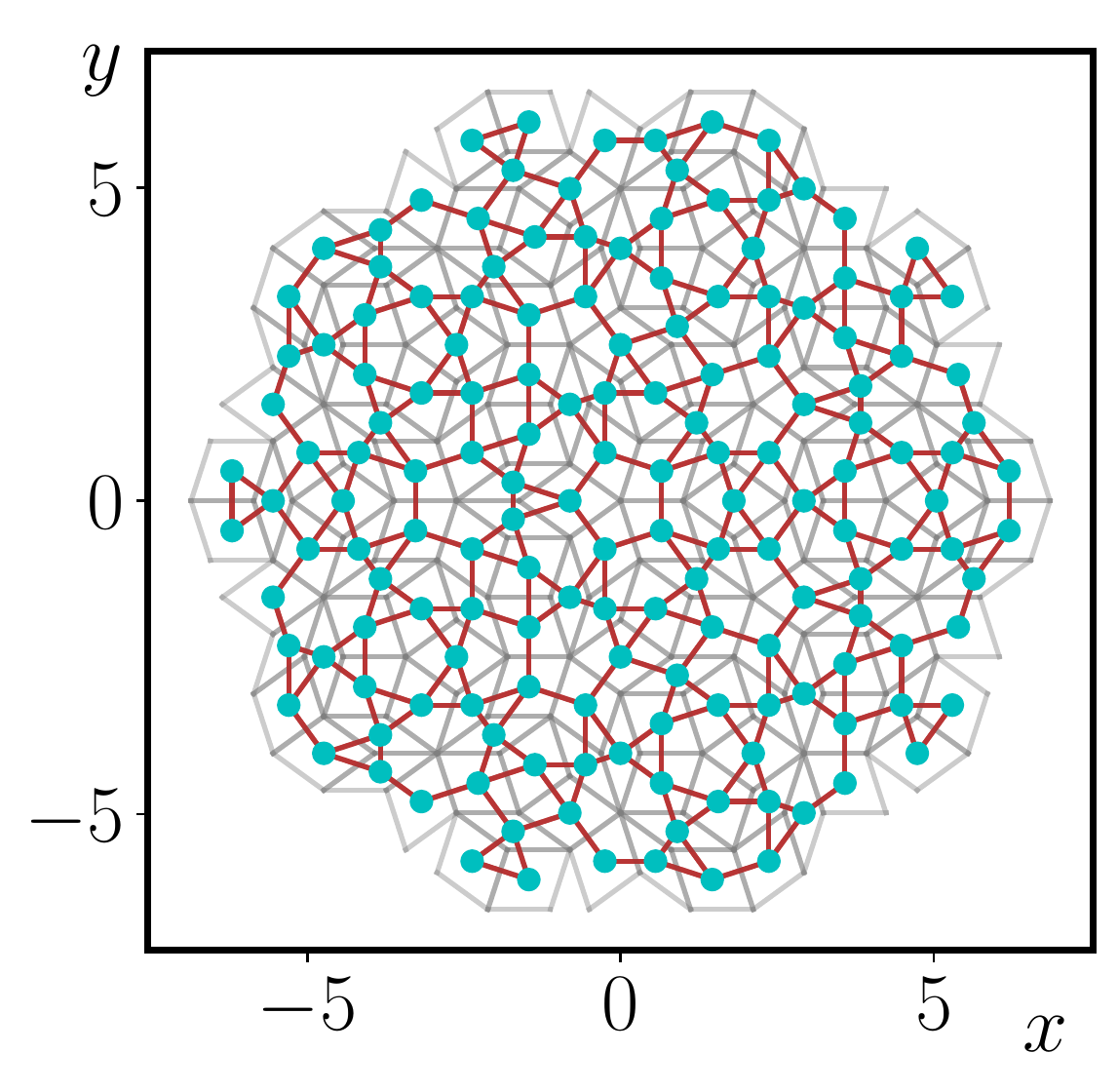}
    \caption{{\bf Illustration of quasicrystal lattices.} Four times iterated Penrose tiling with the vertex model left and the center model right. Nearest neighbors are marked as red bonds.  The rhombi referred to in the main text are shown in grey for the center model. The lattice sites are marked as blue dots. The hoppings in the Hamiltonian are set to $t_0$ only if a red line connects the two sites in question.} 
\label{Fig::hoppings}
\end{figure}

\emph{Model and method. --- }
We examine a Penrose tiling generated by the substitution method~\cite{goodman-strauss_matching_1998} using $10$ triangles as initial configuration.
 Based on this tiling, we construct Hubbard models with sites located on either the vertices or the centers of the rhombi of the lattice, called vertex or center models, respectively (compare Fig.~\ref{Fig::hoppings}). 
 In all simulations presented here, we employ 3126 lattice sites in the vertex model or 3010 in the center model. The phases and critical scales do not change significantly upon further increasing the lattice size. The Hubbard Hamiltonian in second quantization reads
 \begin{align}
 \centering
    \nonumber
     H = -\sum_{\rm i,j, \sigma} (t_{\rm i,j} &+ \mu \delta_{\rm i,j})c^{\dagger}_{\rm i,\sigma} c_{\rm j,\sigma} + \frac{1}{2}\sum_{\rm i,\sigma,\sigma'} U n_{\rm i, \sigma} n_{\rm i, \sigma'} \\
     &+ \frac{1}{2} \sum_{\rm \left\langle i,j\right\rangle,\sigma,\sigma'} U' n_{\rm i, \sigma} n_{\rm j, \sigma'}\;,\label{eq:H}
 \end{align}
  with the operators $c_{\rm i,\sigma}^{(\dagger )}$ annihilating (creating) an electron on site $i$ with spin $\sigma$.
 We concentrate on two cases for the hopping amplitudes $t_{\rm i,j}$. First, we apply nearest-neighbor hoppings for which $t_{\rm i,j}=t_0$ for neighboring sites, shown as red lines in Fig.~\ref{Fig::hoppings}, while for all non-nearest neighbors $t_{\rm i,j}=0$. Second, we consider exponentially decaying hoppings using an exponential form $t_{\rm i,j} = t_{\rm 0} e^{1-\frac{|r_{\rm i}-r_{\rm j}|}{a}}$. Here, $r_{\rm i}$ is the real-space position of the site $i$. We choose the minimal distance of any two sites, $a=\min_{\rm ij} |r_{\rm i}-r_{\rm j}|$, as lattice spacing and measure bond lengths relative to it~\cite{cao_kohn-luttinger_2020}. For convenience, we set units  by $t_{\rm 0}=1$.  
We assume spin-independence of the interaction with an on-site repulsion $U$ and a nearest-neighbor repulsion $U'$, resulting in an $\mathrm{SU(2)}$-symmetric Hamiltonian. This is a convenient, but in no means necessary, simplification for our approach (see Appendix~\ref{App::A}). All of our results are calculated at temperature $T =10^{-3}$.

To treat possible competing orders in an unbiased fashion, we employ a real-space variant of the TUFRG~\cite{lichtenstein_high-performance_2017, bauer_functional_2014, weidinger_functional_2017, markhof_detecting_2018,eckhardt_truncated_2020}, based on an one-loop formulation of the FRG~\cite{metzner_functional_2012, salmhofer_fermionic_2001}. In a FRG scheme, a cutoff function $R(\Lambda)$ is introduced in the bare propagator, such that the system reduces to a solvable problem at an initial scale. At a final scale, the full solution is recovered. By variations of the cutoff parameter $\Lambda$, one obtains an infinite set of differential flow equations.
This set of flow equations needs to be truncated in order to be numerically tractable, lending  a perturbative motivation to the FRG. Here, we employ one of the most commonly applied truncations, keeping only the effective two-particle interaction without its frequency dependence, which was successfully applied before in 2D for the study of competing instabilities in crystalline structures \cite{metzner_functional_2012, honerkamp_interaction_2004, platt_functional_2013, klebl_functional_2020}. 
The diagrams for the effective interaction can be classified in three groups or channels. Each of the channels is related to a separate effective MF Hamiltonian~\cite{metzner_functional_2012}. Thus, a divergence (also called flow to strong coupling) in a certain channel can be directly associated with an emergent order parameter and possibly to a gap opening in the corresponding MF picture, indicative of a phase transition. The emergent phase and its spatial structure are encoded in the type and value of the diverging channel. More specifically, the so-called pairing-channel indicates an opening of the pairing gap, the spin-channel of a magnetic gap  and the charge-channel of a charge gap (see Appendix~\ref{App::B} for details).

{
The real-space TUFRG approach exploits the dependency of each of these channels on so-called "native" indices. These native indices are the dependencies generated in a random phase approximation calculation, which is equivalent to a FRG flow without intra-channel coupling. Dependencies beyond these native indices are only generated at higher orders in the interaction, where the further apart the third and fourth index from the native ones are, the higher the necessary interaction order to generate these contributions gets. Motivated by this, we introduce projections onto each channels' native indices as follows
\begin{widetext}
\begin{align}
\begin{split}
\hat{P}[\Gamma]_{i,j}^{b_i,b_j}  &= \Gamma(i,i+b_i;j,j+b_j) =  \sum_{k,l} \Gamma(i,k;j,l)f_{b_i}(k)f^{*}_{b_j}(l), \\ 
\hat{C}[\Gamma]_{i,j}^{b_i,b_j}  &= \Gamma(i,j+b_j;j,i+b_i) =\sum_{k,l} \Gamma(i,k;j,l)f_{b_i}(l)f^{*}_{b_j}(k), \\ 
\hat{D}[\Gamma]_{i,j}^{b_i,b_j}  &= \Gamma(i,j+b_j;i+b_i,j) =\sum_{k,l} \Gamma(i,k;l,j)f_{b_i}(l)f^{*}_{b_j}(k), 
\end{split}
\label{Eq::projectno_su2}
\end{align}
\end{widetext}
where $\Gamma$ is a general vertex object and the different projections are marked with a letter to associate them to their respective channel. The pairing-channel is abbreviated as $P$, the spin-channel as $C$ and the charge-channel as $D$. The form-factors or bonds $f_{b_l}(l)$ form an orthonormal basis on the lattice defined by
\begin{align}
\begin{split}
\sum_{i} f_{b_i}(i)  f^{*}_{b_i'} (i) &= \delta_{b_i,b_{i'}},\\
\sum_{b_k} f_{b_k}(i)f^{*}_{b_i}(i') &= \delta_{i,i'}.
\end{split}
\label{Eq_Formfactor}
\end{align}

Due to the pertubatively motivated character of FRG it is reasonable to neglect terms generated at high orders in the interaction, which translates to restricting the bonds used in the expansions in Eq.~\ref{Eq::projectno_su2} to a small subset of the full lattice, effectively reducing the size of the projected channels. The flow equations for the projected channels are obtained by inserting the definitions in the original flow equations and introducing a productive unity, see Appendix~\ref{App::A} for a detailed derivation. Leading to the following flow equations in the $SU(2)$ invariant case
\begin{align}
\begin{split}
    \frac{dP}{d\Lambda}  &= -\hat{P}[\Gamma] \cdot \dot{\chi}_{pp} \cdot \hat{P}[\Gamma], \\
    \frac{dC}{d\Lambda} &= -\hat{C}[\Gamma] \cdot \dot{\chi}_{ph} \cdot \hat{C}[\Gamma], \\
    \frac{dD}{d\Lambda} &= 2\hat{D}[\Gamma] \cdot \dot{\chi}_{ph} \cdot \hat{D}[\Gamma]\\ &- \hat{C}[\Gamma] \cdot \dot{\chi}_{ph} \cdot \hat{D}[\Gamma] \\ &-\hat{D}[\Gamma] \cdot \dot{\chi}_{ph} \cdot \hat{C}[\Gamma],
\end{split}
\label{equ:flow_equations}
\end{align}
where $P$, $C$ and $D$ are the respective channels projected on their main dependencies and $\Gamma$ is the effective interaction reconstructed by
\begin{equation}
\Gamma = U + \hat{P}^{-1}[P] + \hat{C}^{-1}[C] +\hat{D}^{-1}[D].
\end{equation}
$\dot{\chi}_{pp}$ and $\dot{\chi}_{ph}$ are the scale differentiated particle-particle and particle-hole propagators which can be calculated from the Greens function $G = R(\lambda)(i\omega-H)^{-1}$ and the single scale propagator $S = \partial_{\Lambda}R(\lambda)(i\omega-H)^{-1}$ by
\begin{align*}
    \dot{\chi}_{ph  (i,j)}^{b_i,b_j} &= 2T\sum_{\omega>0} \Re{\left(G(\omega)_{i,j}S(\omega)_{j+b_j,i+b_i} + G \leftrightarrow S\right)}, \\
    \dot{\chi}_{pp  (i,j)}^{b_i,b_j} &= 2T\sum_{\omega>0} \Re{\left(G(\omega)_{i,j}S(-\omega)_{i+b_i,j+b_j} + G \leftrightarrow S\right)},
\end{align*}
where $T$ is the temperature.
In this formula we already used the symmetry of the summand w.r.t.~frequency to reduce the numerical effort. Throughout this paper we will use the so called $\Omega$-cutoff~\cite{husemann_efficient_2009} given by
\begin{equation}
    R(\Lambda) = \frac{\omega^2}{\omega^2+\Lambda^2}.
\end{equation}
}
 In practice, the spatial ordering is extracted from the leading eigenvectors of the diverging channel. We stop the flow if an eigenvalue of one of the three channels surpasses a threshold corresponding to the stopping scale denoted as $\Lambda = \Lambda_c$.

Due to the employed open boundary conditions the kinetic energy is reduced at the edges of the lattice, increasing the relative relevance of the interaction there. To avoid phases arising solely due to this boundary effect, in case they do not coexist with a spatially separated bulk phase, we employ a $\tanh$ envelope falling off to the boundary:
\begin{align}
    U_{ij}^{sc}\nonumber =\frac{U_{ij}}{2}\cdot&[\tanh(10(d_{\rm max}-d_i)^2)+\\&\tanh(10(d_{\rm max}-d_j)^2)],
\label{Eq:sym_scalapp}
\end{align}
where $d_i = |r_i|$ and $d_{\rm max} = {\rm max}_i d_i$. 
Introducing this envelop, we can  probe the bulk phase-diagram more easily which is related to the thermodynamic limit by the self-similarity of the lattice.

\emph{Competing orders in the vertex model. --- }
First, we investigate the vertex model including nearest-neighbor hopping at half filling ($\mu = 0.0$). The model is bipartite and known to show antiferromagnetic ordering in the case of $U'=0$ and $U>0$~\cite{koga_antiferromagnetic_2017, jagannathan_penrose_2007}. The density of states (DoS) has a $\delta$-like peak at $\omega = 0.0$ separated by a small gap from the rest of the spectrum (see Appendix~\ref{App::D}). This peak does not arise due to a Van-Hove singularity (which are very relevant in the context of ordering in crystals), but instead occurs due to a macroscopic number of degenerate states with eigenvalue zero \cite{kohmoto_electronic_1986-1,kohmoto_electronic_1986,day-roberts_nature_2020}. To access the bulk phase-diagram of the vertex model we employ the envelope Eq.~\eqref{Eq:sym_scal}.
 
\begin{figure}
\centering
{\includegraphics[width = 1\linewidth]{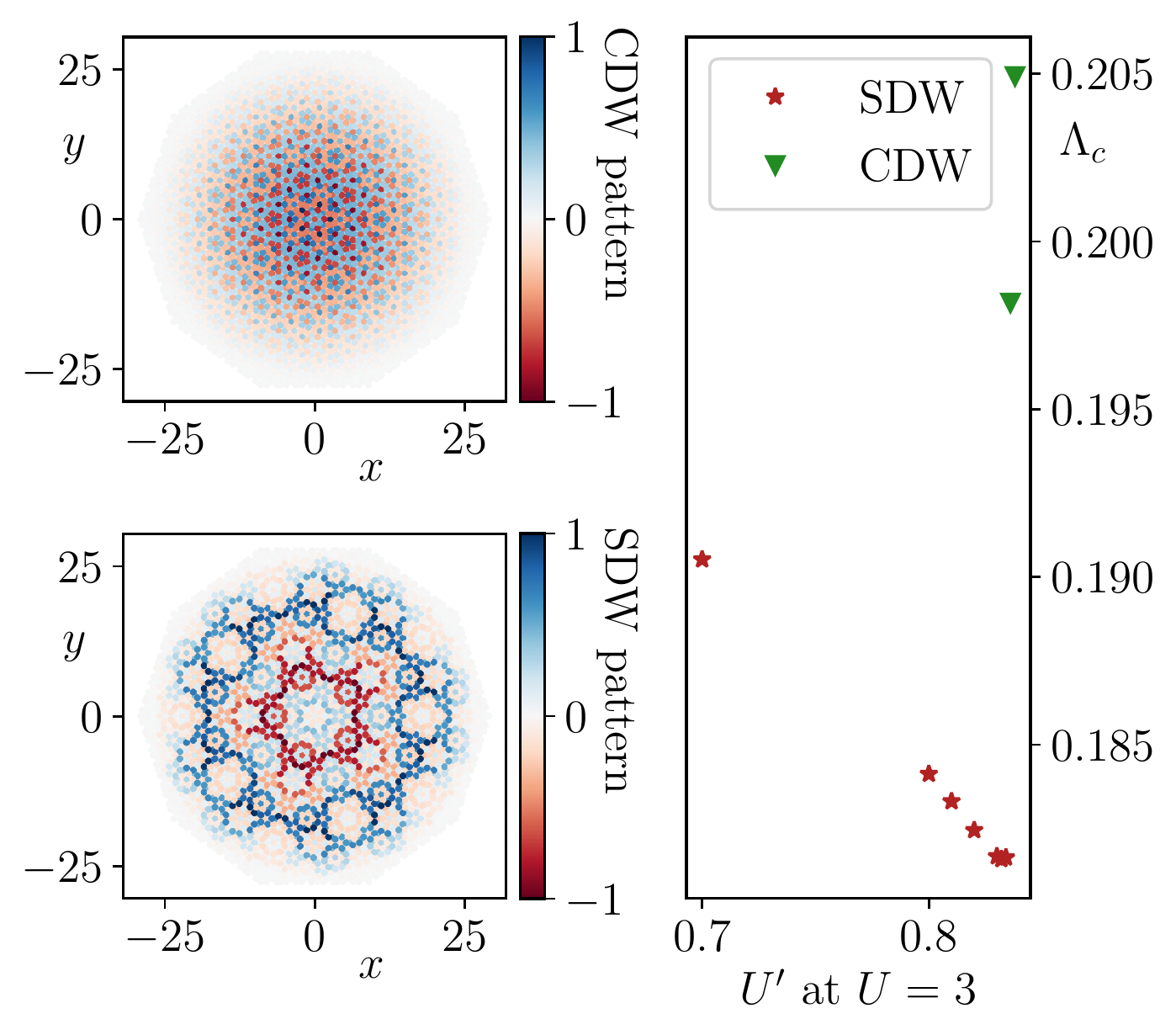}}
\caption{{\bf Vertex model at $\mu = 0.0$: Competition of order}. We employing Eq.~\eqref{Eq:sym_scalapp} for the interaction. The upper left plot visualizes the charge order parameter at $U'=1.5$ and $U=1$. The lower left plot visualizes the spin order parameter (or magnetization pattern), at $U'=0.5$ and $U=2.0$. On the right side, the critical scales depending on the nearest-neighbor interaction are shown for $U = 3.0$ in the vicinity of the phase transition, to highlight the suppression of the critical scale upon approaching the phase transition.  We find a competition of the two phases at the transition, similar to the behavior prototypically found in conventional crystals.}
\label{Fig::screened}
\end{figure}

We encounter two different phases. For $U\gg U'$, an antiferrromagnetic spin-density wave (SDW) instability prevails with a similar ordering pattern (see lower left plot in Fig.~\ref{Fig::screened}) to the ordering pattern at half-filling without nearest-neighbor interactions ($U'=0$)~\cite{koga_antiferromagnetic_2017}. For increased $U'/U$, the leading divergence changes to a charge-density wave (CDW) (see upper left plot in Fig.~\ref{Fig::screened}). {We do not discuss these orderings in detail here, as on the one hand such discussions can be found elsewhere~\cite{kohmoto_electronic_1986-1,jagannathan_penrose_2007} and on the other hand we want to focus on the interplay in between the two ordering types.}
The transition between SDW and CDW is accompanied by a reduction of the critical scales, as can be seen in the right plot in Fig.~\ref{Fig::screened}. The critical scale without coupling the different channels with each other is $\Lambda_c\approx 0.4$ at the transition, but the ordering vectors are predicted correctly by this simplified calculation. This implies that the divergences are generated within each of the channels individually, in contrast to fluctuation driven divergences which are generated by the feedback of one channel to another. In the simulation incorporating inter-channel feedback, the critical scale is reduced to $\Lambda_c\approx 0.18$. Thus, we conclude that the ordering tendencies compete, leading to a mutual suppression of the phases.

To sum up, the physics of the {\it vertex} model is similar to the one of crystals with translational invariance, where multiple mutually competing bulk orders take center stage.

\emph{Spatial coexistence of order in the nearest neighbor center model. --- }
Next, we examine a center model including nearest-neighbor hopping and interactions without employing Eq.~\eqref{Eq:sym_scal}, as here a bulk-boundary coexistence of order is observed. This model is not bipartite and, in contrast to the vertex model, each site has the coordination number four. Its DoS (see Appendix~\ref{App::D}) displays two main peaks consisting of a macroscopic number of degenerate states (as discussed above): One at $\omega\approx 2.35$ and one at $\omega = 2.00$. 

We concentrate on a Fermi level at the main features at $\omega = 2.00$, which arises in parts due to so-called string states, which are self-similar states with a fractional dimension of $1.44$ \cite{fujiwara_localized_1988, inayoshi_excitonic_2020}. Here, we find a CDW as well as two different SDW divergences depending on the values of $U$ and $U'$. At $U \gg U'$ we find an on-site SDW ordering, whereas at $U' \gg U$ we find an on-site CDW ordering (see Appendix~\ref{App::F}). In the transition region between the two, a SDW phase with a more complex ordering pattern emerges. This leads to two distinct phase transitions, one is a smooth interpolation from a bulk ordered on-site SDW to a boundary pinned phase, the second one is a transition between the latter phase and a bulk ordered on-site CDW. The second transition has a mutually evasive nature, as the two occurring orderings have vanishing spatial overlap as we will analyze next. 
The region in which the transition between SDW and CDW occurs is of special interest to us. In contrast to the vertex model, the strongest two channels, namely charge and spin channel, diverge on equal footing and have main weight in separate regions. 
This separation is leading to a gap opening in different spatial regions of the lattice, i.e., a coexistence of two different orderings. To show this more clearly, we proceed with a MF decoupling of the effective FRG interaction at the final scale $\Lambda_c$ (see Appendix~\ref{App::B} or \cite{reiss_renormalized_2007}).
\begin{figure}
       \includegraphics[width = 1\columnwidth]{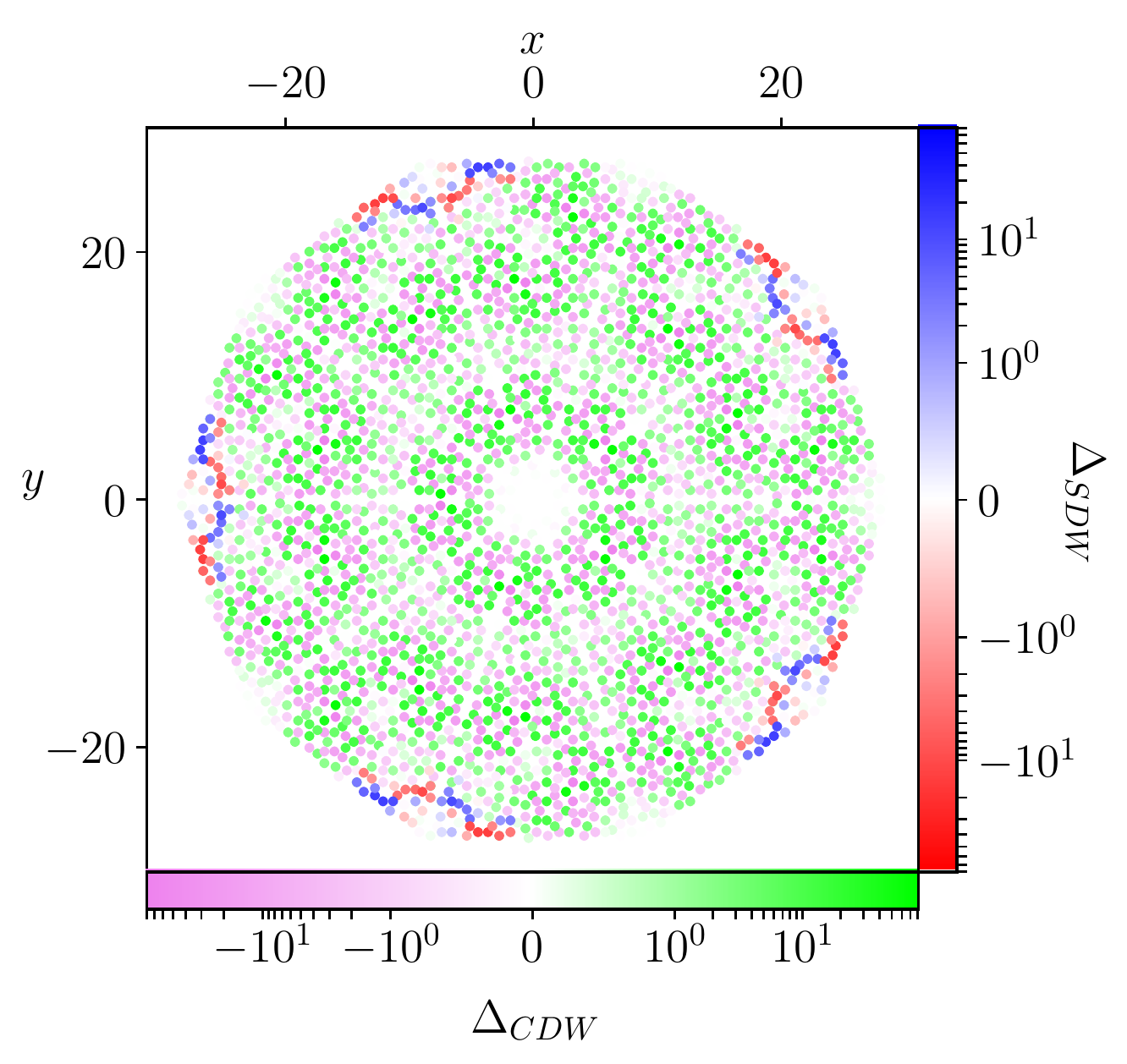}
\caption{{\bf Center model at $\mu = 2.0$: Coexistence of order}. Gap values of the decoupled effective interaction at $\Lambda_c$ for $U = 1$ and  $U'=1.015$. Leading ordering strengths of the charge and spin ordering channels have approximately the same magnitude. The calculated gap magnitudes are shown in a red-blue color scheme for the spin gap, which is equivalent to the magnetisation. In the violet-lime color scheme the charge gap is shown. The two faces have no spatial overlap of magnitudes above $10^{-3}{\rm max}(\Delta)$. The coexistence of the two phases is found to be a general feature of this phase transition.}
\label{Fig::coexistence_of_phases}
\end{figure}
The spatially separated support of the non-zero gaps for charge and spin order are shown in Fig.~\ref{Fig::coexistence_of_phases}. The charge order forms in the center as a self-similar structure and shows alternating signs.
The spin order emerges close to the boundary of the system and has strongly peaked maximal values separated by an order of magnitude from the lower gap values. In the transition region separating the phases both have low weight. This behavior arises generically for $U$ and $U'$ combinations at the transition line between the charge and spin ordering. The coexistence is quite sensitive to changes in $U$ and $U'$, as moving away from the transition line rapidly promotes a single phase. Due to the spatially separated nature, the two orders coexist without affecting each other. 

Summarizing these findings, we have identified a coexistence of mutually evasive orders with clear spatial separation in the {\it center} Penrose model. Such instabilities can likely only be found in structures with (infinitely) large unit cells such as true quasicrystals, their large-scale approximants, or twisted moir\'e materials. 

\emph{Collaborative order in the center model with exponentially decaying hopping amplitudes. --- }
Finally, we investigate the phase-diagram of a center model with hopping amplitudes decaying exponentially in real-space. For such models, recently an unconventional superconducting order has been predicted~\cite{cao_kohn-luttinger_2020}. 
The DoS in this setup has three main peaks at $\omega \approx 0.83,\,0.99,\,1.23$ (see Appendix~\ref{App::D}). Between the second and the third peak, there is a small energy gap. The most interesting physics is again expected for the Fermi level at the DoS peaks or in their vicinity.
For the simulations we chose $U'=0$ and again employ Eq.~\eqref{Eq:sym_scal} in order to suppress the boundary states.

\begin{figure}[!htbp]
    \includegraphics[width = 0.49\columnwidth, height = 0.388\columnwidth]{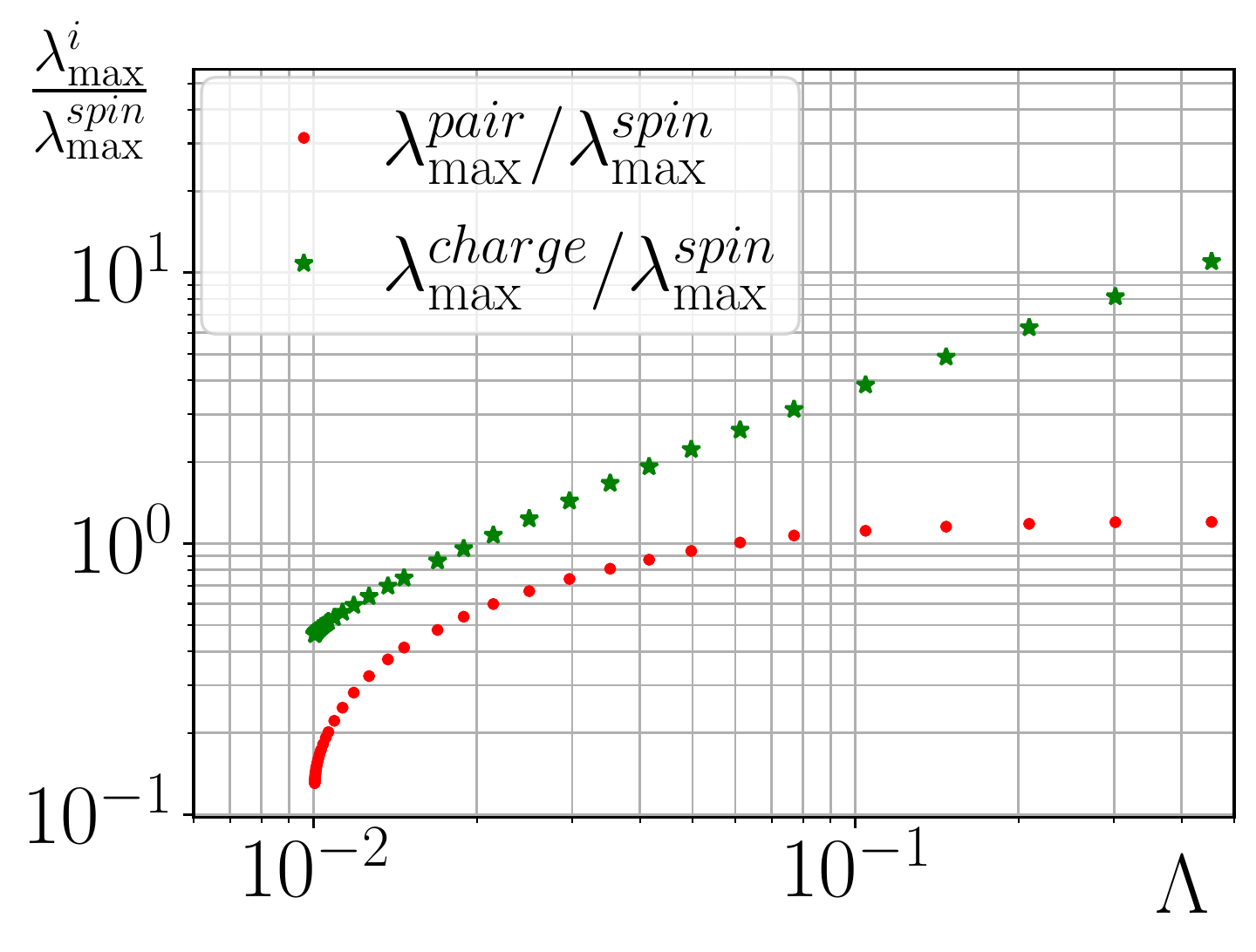}
    \includegraphics[width = 0.49\columnwidth]{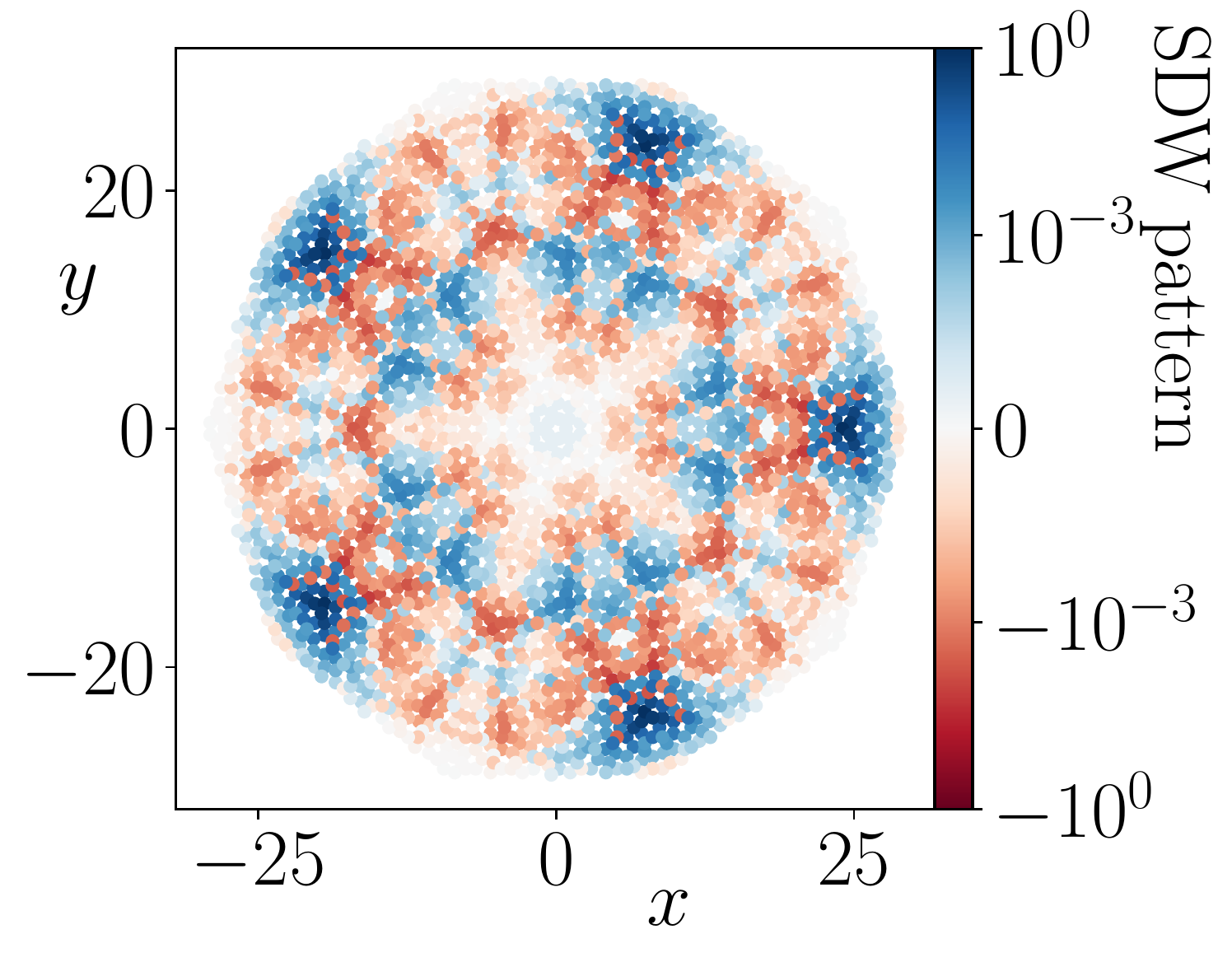}
    \vfill
    \includegraphics[width = 0.49\columnwidth]{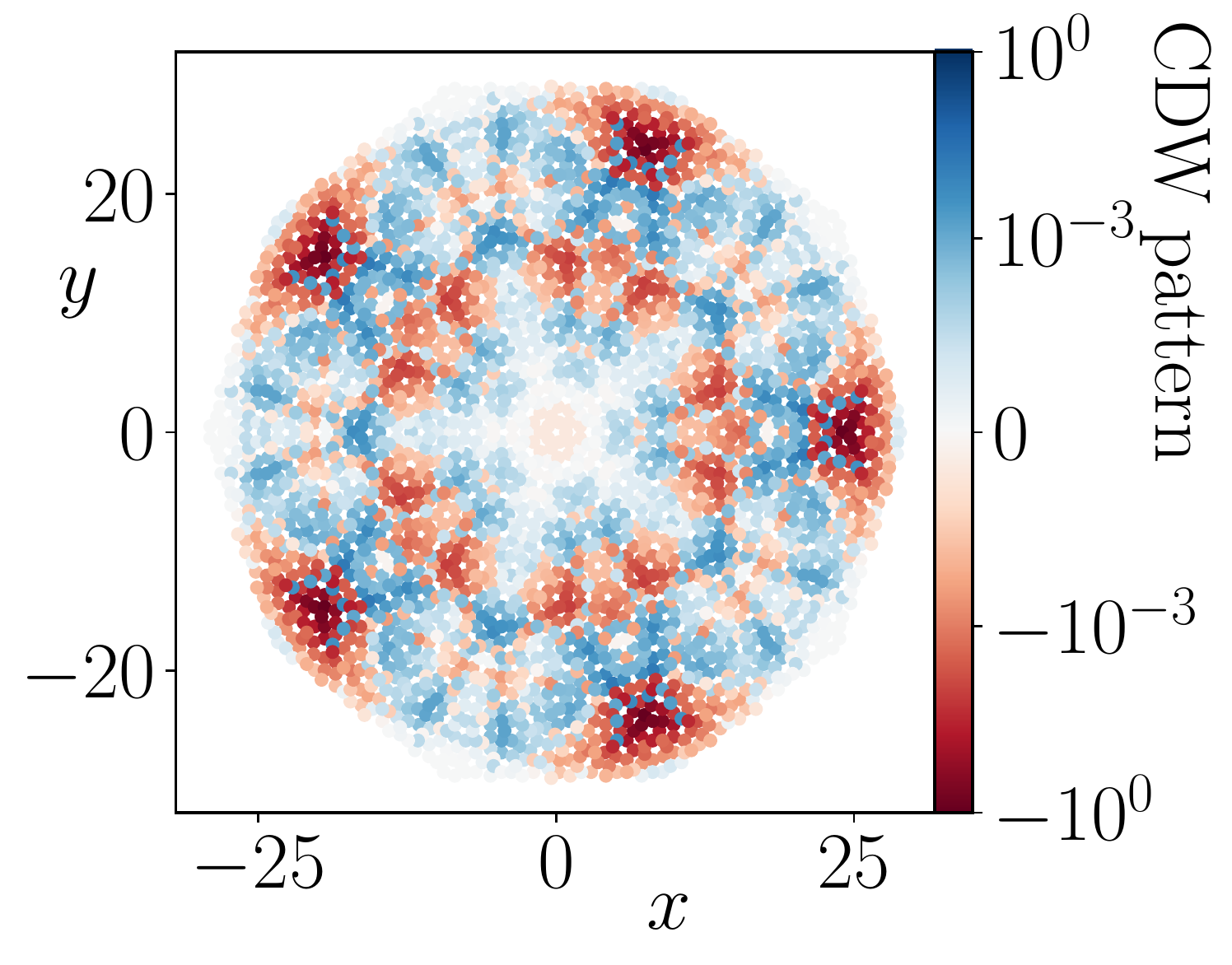}
       \includegraphics[width = 0.49\columnwidth]{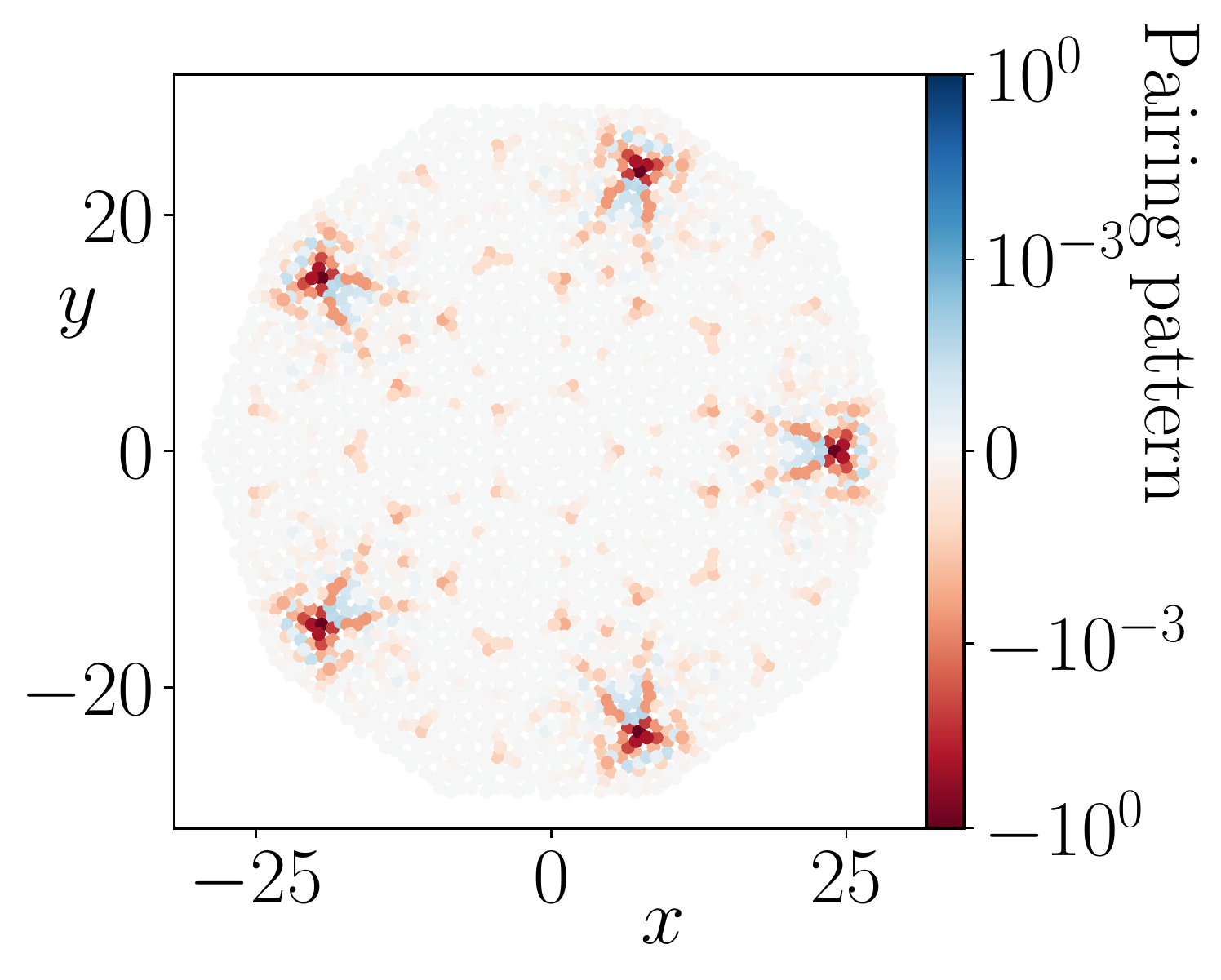}
\caption{{\bf Center model with exponentially decaying hopping amplitudes at $\mu = 0.99$: Collaboration of order.} 
In the upper left, the maximal eigenvalues of the pairing- and charge-channel relative to the one of the spin-channel are shown for $U = 0.3$. The expected ordering pattern for the spin-channel is visualized in the upper right plot, the one for the charge-channel in the lower left plot and the one for the pairing channel in the lower right plot. We observe a slow decay of the relative maximal eigenvalues pointing at a possible unconventional groundstate. The ordering patterns of the three channels have spatial overlap, therefore there is no spatial evasion of order.
}

\label{Fig::all_diverge}
\end{figure}
A SDW is the dominant phase in the whole interaction region at all three peaks of the DoS. Its ordering pattern is dependent on the filling as well as the interaction strength $U$ (see Appendix~\ref{App:G}). Upon decreasing $U$ to $0.3$ we observe the emergence of a sub-leading pairing divergence indicated by a slower decay of the fraction of maximal eigenvalues of the pairing- and spin-channel $\nicefrac{\lambda^{pair}_{max}}{\lambda^{spin}_{max}}$ than at higher interactions. This is accompanied by the loss of the standard  behavior for an emergent spin ordering, namely that $\nicefrac{\lambda^{charge}_{max}}{\lambda^{spin}_{max}} = \nicefrac{1}{2}$ \cite{metzner_functional_2012}. Instead we find that $\nicefrac{\lambda^{charge}_{max}}{\lambda^{spin}_{max}}$ decreases slightly upon approaching the critical scale, see Fig.~\ref{Fig::all_diverge}. These findings imply only slightly different critical exponents for the three different ordering tendencies. 
Analogous multi-channel instabilities are known to signal non-trivial non-MF ground states in one-dimensional models for correlated fermions \cite{fisher_mott_1998, lin_exact_1998,hur_superconductivity_2009} and were argued to indicate Fermi surface truncations in the two-dimensional Hubbard models \cite{furukawa_truncation_1998, lauchli_d_2004,honerkamp_breakdown_2001}. Yet, in this case as well as in ours order collaborate and a proper classification of the true ground state needs further work. In any case, it is exciting to see that quasicrystalline systems offer another playground to exhibit such physics at the frontier of our understanding.

To sum up, in the center model with {\it exponentially decaying hopping amplitudes} we find a delicate mutual collaboration of ordering tendencies, pointing to an unconventional ground state. Here MF decouplings are highly biased and a more sophisticated approach needs to be employed. Recent theoretical reports of superconductivity should therefore be taken with caution if they focus on one particular channel in a MF treatment~\cite{cao_kohn-luttinger_2020}.

\emph{Conclusions. --- }
We used a newly developed real-space TUFRG formalism to study the electronic instabilities in quasicrystals. With their infinite size unit cells, the spatial degree of freedom in these systems opens up even greater variety of ordering tendencies beyond competition, allowing for spatial coexistence by mutual evasion and joint, collaborative instabilities.
We expect similar findings for large but finite unit cell systems, with twisted van der Waals heterostructures as most prominent \cite{moiresim} but certainly not last example.

An intriguing avenue of future research concerns the study of disordered vertex arrangements or quasicrystalline approximants which could help in understanding the interesting interplay between interactions and lattice geometry in quasicrystals. {Additionally, an in depth study of the observed orderingss robustness towards rotational symmetry breaking of the lattice could reveal new emerging orderings.}
As a next step, it is desirable to keep a higher accuracy, e.g.~by taking into account the self-energy effects \cite{hille_quantitative_2020} and the frequency dependence of the vertex \cite{Wentzell_2020,weidinger_functional_2017, reckling_approximating_2018,vilardi_nonseparable_2017}. 
This allows to include interactions mediated by phonons or photons and thus to study conventional superconductivity in quasicrystals~\cite{araujo_conventional_2019}. On the level of real materials the next step is the development of a feasible beyond inter-orbital-bilinears \cite{honerkamp_efficient_2018} approach, combining the here employed real-space TUFRG with the momentum-space TUFRG~\cite{lichtenstein_high-performance_2017}. With such a scheme $3$D-quasicrystalline approximants can be addressed, opening the ground for predictions of phase-diagrams of real materials if combined with density functional theory inputs~\cite{hohenberg_inhomogeneous_1964,kohn_self-consistent_1965}.

\begin{acknowledgements}
We thank J.\ Ehrlich, L.\ Klebl and J.\ Beyer for fruitful discussions and D.\ Rohe and E. Di Napoli for the help in optimizing the framework. The Deutsche Forschungsgemeinschaft (DFG, German Research Foundation) is acknowledged for support through RTG 1995 and under Germany's Excellence Strategy - Cluster of Excellence Matter and Light for Quantum Computing (ML4Q) EXC 2004/1 - 390534769. We acknowledge support from the Max Planck-New York City Center for Non-Equilibrium Quantum Phenomena.
Simulations were performed with computing resources granted by RWTH Aachen University under project rwth0545.
\end{acknowledgements}

\bibliography{2ndrev_MAIN}
\clearpage
\onecolumngrid
\appendix

\section{Derivation of the flow equations}
\label{App::A}
We now sketch the derivation of the flow equations.
The effective vertex can be separated in three different channels, each of them related to a specific fermionic bilinear \cite{metzner_functional_2012}. A divergence of their eigenvalues indicates a flow to strong coupling, which in turn can lead to the opening of a specific gap in the self-energy. The $P$-channel indicates pairing, the $C$-channel gives information of magnetic ordering tendencies, and charge ordering information is contained in the $D$-channel. We called the $P$-channel pairing-channel, the $C$-channel spin-channel and the $D$-channel charge-channel in the main body to highlight their physical meaning. Technically the charge ordering information has to be extracted from the physical charge channel~\cite{husemann_efficient_2009}, but if no other channels diverge, the ordering can be extracted directly from the $D$-channel or as called in the main body the charge-channel. This can be seen by a mean-field decoupling in the three native channels, see below.
In general, we can write the effective two-particle interaction $\Gamma^{(4)}$ as (suppressing the dependency on the flow parameter $\Lambda$)
\begin{equation}
\Gamma^{(4)}(1,2;3,4) = U(1,2;3,4)+\overwrite{\Phi^P(1,2;3,4)}{green}{Particle-Particle (PP)} +\overwrite{\Phi^D(1,2;3,4)}{red}{Direct Particle-Hole (PH)} +\overwrite{\Phi^C(1,2;3,4)}{orange}{Exchange/ Crossed PH}. 
\end{equation}
Each number represents a multi-index consisting of orbital, spin and frequency index. We will use quoted numbers, or quoted indices, for all degrees of freedom which are summed.
The flow equation for the effective Interaction can be separated in those three channels~\cite{lichtenstein_high-performance_2017} as can be seen in Eq.~\eqref{Eq::gamma4P}-\eqref{Eq::gamma4D}. As all occurring quantities are $\Lambda$-dependent we will from now on leave out the subscript.
\begin{subequations} 
\begin{align}
\frac{d\Phi^P(1,2;3,4)}{d\Lambda} = -\sum \Gamma(1,2;1',2')&\left(G(1',3')S(2',4') \right) \Gamma(3',4';3,4), \label{Eq::gamma4P}\\
\nonumber \frac{d\Phi^C(1,2;3,4)}{d\Lambda} = \sum  \Gamma(1,4';1',4)&(G(1',3')S(2',4') \\+ &S(1',3')G(2',4')) \Gamma(3',2;3, 2'),\label{Eq::gamma4C}  \\
\nonumber \frac{d\Phi^D(1,2;3,4)}{d\Lambda} = -\sum  \Gamma(1,3';3,2')&(G(1',3')S(2',4') \\+ &S(1',3')G(2',4')) \Gamma(4',2;1',4). 
\label{Eq::gamma4D}
\end{align}
\end{subequations}
We will neglect the frequency dependence of the vertex, therefore we switch to letters as indices, each of which describes a lattice point with associated spin. 
The truncated unity approach we develop is analog to the momentum-space one. There, the main idea is that without inter-channel coupling the equations amount to a standard RPA series which will only produce a dependency on two spatial indices, or a single momentum (if the interaction is local). If we now include the coupling, we will technically generate dependencies on all indices, but the "native" ones will contain the main features, as configurations beyond those are only generated at higher order in the interaction strength. This indicates that only two orbital indices are of central importance and should be treated with high accuracy and for the others a lower accuracy is sufficient. We want to exploit this by an expansion of each non-native indices in a basis centered at a native one. In a general real-space setting the basis is dependent on the position it is centered around. Mathematically speaking we use a form-factor expansion of pairwise orthonormal functions which form a complete basis (see Eq. \eqref{Eq_Formfactor_app}):
\begin{align}
\begin{split}
\sum_{i} f_{b_k}(i)  f^{*}_{b_k'} (i) &= \delta_{b_k,b_{k'}},\\
\sum_{b_k} f_{b_k}(i)f^{*}_{b_k}(i') &= \delta_{i,i'}.
\end{split}
\label{Eq_Formfactor_app}
\end{align}
Later the included bonds in the unity are restricted to a small subset.
We start with general bonds for the derivation and specify them later on. With the usage of a basis we define the projections onto the main dependencies of each channels in Eq.~\eqref{Eq::projectno_su2_app} as
\begin{align}
\begin{split}
\hat{P}[\Gamma]_{i,j}^{b_i,b_j}  &= \Gamma(i,i+b_i;j,j+b_j) =  \sum_{k,l} \Gamma(i,k;j,l)f_{b_i}(k)f^{*}_{b_j}(l), \\ 
\hat{C}[\Gamma]_{i,j}^{b_i,b_j}  &= \Gamma(i,j+b_j;j,i+b_i) =\sum_{k,l} \Gamma(i,k;j,l)f_{b_i}(l)f^{*}_{b_j}(k), \\ 
\hat{D}[\Gamma]_{i,j}^{b_i,b_j}  &= \Gamma(i,j+b_j;i+b_i,j) =\sum_{k,l} \Gamma(i,k;l,j)f_{b_i}(l)f^{*}_{b_j}(k). 
\end{split}
\label{Eq::projectno_su2_app}
\end{align}
This is exact and just a rewriting of the vertex as long as we do not truncate the basis. The full vertex can be recovered by a unprojection, the inverse projection, using the completeness relation in Eq.~\eqref{Eq_Formfactor}. Further we define the channels projected on their respective native indices in Eq.~\eqref{Eq::project}.
\begin{align}
\begin{split}
P_{i,j}^{b_i,b_j} &= \hat{P}[\Phi^P]_{i,j}^{b_i,b_j}  \\ 
C_{i,j}^{b_i,b_j} &= \hat{C}[\Phi^C]_{i,j}^{b_i,b_j} \\ 
D_{i,j}^{b_i,b_j} &= \hat{D}[\Phi^D]_{i,j}^{b_i,b_j}
\end{split}
\label{Eq::project}
\end{align}
Thus we recover the full vertex by applying
\begin{equation}
\Gamma = U + \hat{P}^{-1}[P] + \hat{C}^{-1}[C] +\hat{D}^{-1}[D].
\end{equation}

The flow equations for the projected channels can be derived from Eq.~\eqref{Eq::gamma4P}-\eqref{Eq::gamma4D} with the help of an insertion of an unity, here shown exemplary for the $P$-channel
\begin{align}
-\frac{dP_{i,j}^{b_i,b_j} }{d\Lambda} &= \sum_{k,l} \frac{d\Phi^P(i,k;j,l) }{d\Lambda}f_{b_i}(k)f^{*}_{b_j}(l) \\ 
&= \nonumber\sum_{k,l,i',k',j',l'} f_{b_i}(k)f^{*}_{b_j}(l)\Gamma(i,k;i',k')\sum_{n',m'} \delta_{n',k'}  \left(G(i',j')S(n',m') \right) \delta_{m',l'}  \Gamma(j',l';j,l) \\
&= \nonumber\sum_{k,l,i',k',j',l'} f_{b_i}(k)f^{*}_{b_j}(l)\Gamma(i,k;i',k')\sum_{b_{i'},n',b_{j'},m'} [f_{b_{i'}}(k')f^{*}_{b_{i'}}(n')  \left(G(i',j')S(n',m') \right) \\  &\quad \times  f_{b_{j'}}(l')f^{*}_{b_{j'}}(m')  \Gamma(j',l';j,l)] \\
&= \nonumber \sum_{b_i',i',j',b_j'} \left[\sum_{k,k'}f_{b_i}(k)f^{*}_{b_{i'}}(k')\Gamma(i,k;i',k')\right]\left[\sum_{n',m'} f_{b_{i'}}(n')f^{*}_{b_{j'}}(m')  \left(G(i',j')S(n',m') \right)\right] \\  &\quad \times \left[\sum_{l',l} f_{b_{j'}}(l')f^{*}_{b_{j}}(l)  \Gamma(j',l';j,l)\right] \\
&= \sum_{b_i',i',j',b_j'} \hat{P}[\Gamma]_{i,i'}^{b_i,b_{i'}} \dot{\chi}_{pp(i',j')}^{b_{i'},b_{j'}}  \hat{P}[\Gamma]_{j',j}^{b_{j'},b_{j}}.
\end{align}
The other equations can be derived analogously, resulting in the following coupled set of differential equations
\begin{align}
\label{Eq::Pproj}
\frac{dP_{i,j}^{b_i,b_j} }{d\Lambda} &= -\hat{P}[\Gamma]_{i,i'}^{b_i,b_{i'}} \dot{\chi}_{pp  (i',j')}^{b_i',b_j'}  \hat{P}[\Gamma]_{j',j}^{b_{j'},b_{j}}\\ 
\label{Eq::Cproj}
\frac{dC_{i,j}^{b_i,b_j} }{d\Lambda} &= \hat{C}[\Gamma]_{i,i'}^{b_i,b_{i'}} \dot{\chi}_{ph  (i',j')}^{b_i',b_j'}  \hat{C}[\Gamma]_{j',j}^{b_{j'},b_{j}} \\ 
\frac{dD_{i,j}^{b_i,b_j} }{d\Lambda} &= -\hat{D}[\Gamma]_{i,i'}^{b_i,b_{i'}} \dot{\chi}_{ph  (i',j')}^{b_i',b_j'}  \hat{D}[\Gamma]_{j',j}^{b_{j'},b_{j}},
\label{Eq::Dproj}
\end{align}
where we defined the particle-particle and particle-hole propagator as
\begin{align}
\dot{\chi}_{pp  (i',j')}^{b_i',b_j'}  &= \sum_{n',m'} f_{b_{i'}}(n')f^{*}_{b_{j'}}(m')  \left(G(i',j')S(n',m') \right) \\
\dot{\chi}_{ph  (i',j')}^{b_i',b_j'}  &= \sum_{n',m'} f_{b_{i'}}(n')f^{*}_{b_{j'}}(m')  \left(G(i',j')S(n',m') + S(i',j')G(n',m') \right).
\end{align}
The equations can be reinterpreted in terms of multi-index blockmatrix-products which are well optimized numerically. 

As already mentioned we now want to truncate the basis expansion used in the derivation. This will in fact not change the projected flow equations, but it will change the vertex reconstruction as the inverse projection is not exact anymore. Therefore we obtain projections in between the three channels. The way how the basis is chosen as well as how it is truncated is not unique. The choice of a specific truncation can either be physically or systematically motivated. For example, the first one can be applied in Moir\'e-lattice models where we have two competing length scales. It is believed that such models can exhibit superconductivity at the scale of unit cell vectors which then could be included explicitly in the expansion. The second approach excludes contributions above a certain bond length or distance in a controlled manner. We can express this truncation as
\begin{equation}
\sum_{b_k \in L} f_{b_k}(i)  f^{*}_{b_k} (i) = \sum_{b_k} \delta_{b_k}^L f_{b_k}(i)  f^{*}_{b_k}(i) \approx 1,
\end{equation}
where we defined the set of all allowed bonds $L$. We define $\delta_{b_k}^L$ to be one if $b_k \in L$ and zero otherwise.
To obtain the explicit projections needed for the flow equations we now insert this into the full vertex projections and keep track of all occurring indices. Without specifying a specific basis or writing out spin dependencies, we obtain the general form of the projections as

\begin{align}
\hat{P}[\Gamma]_{i,j}^{b_i,b_j} &= \hat{P}\left[U + \hat{P}^{-1}[P] + \hat{C}^{-1}[C] +\hat{D}^{-1}[D]\right]_{i,j}^{b_i,b_j}, \\
\hat{C}[\Gamma]_{i,j}^{b_i,b_j} &= \hat{C}\left[U + \hat{P}^{-1}[P] + \hat{C}^{-1}[C] +\hat{D}^{-1}[D]\right]_{i,j}^{b_i,b_j},\\
\hat{D}[\Gamma]_{i,j}^{b_i,b_j} &= \hat{D}\left[U + \hat{P}^{-1}[P] + \hat{C}^{-1}[C] +\hat{D}^{-1}[D]\right]_{i,j}^{b_i,b_j}. 
\end{align}
Before specifying our basis we take care of the spin degrees of freedom.
 
\subsection{$SU(2)$ symmetric formulation}
\label{Sec::sym_form}
To get rid off the spin degrees of freedom we will now assume that we have a $SU(2)$-symmetric interaction and Hamiltonian, which allows to simplify the channels with the help of the crossing relations~\cite{salmhofer_fermionic_2001}
\begin{align}
    \nonumber\Phi^P(ik;jl)_{\sigma_i\sigma_k\sigma_j\sigma_l} &= V^P(ki;jl)\delta_{\sigma_i\sigma_j}\delta_{\sigma_k\sigma_l}-V^P(ik;jl)\delta_{\sigma_i\sigma_l}\delta_{\sigma_k\sigma_j}\\
    \Phi^C(ik;jl)_{\sigma_i\sigma_k\sigma_j\sigma_l} &= V^D(ki;jl)\delta_{\sigma_i\sigma_j}\delta_{\sigma_k\sigma_l} -V^C(ik;jl)\delta_{\sigma_i\sigma_l}\delta_{\sigma_k\sigma_j}  \label{Eq::spin_elimination}\\
    \nonumber\Phi^D(ik;jl)_{\sigma_i\sigma_k\sigma_j\sigma_l} &= V^C(ki;jl)\delta_{\sigma_i\sigma_j}\delta_{\sigma_k\sigma_l} -V^D(ik;jl)\delta_{\sigma_i\sigma_l}\delta_{\sigma_k\sigma_j}.
\end{align}
The spin degrees of freedom can now be eliminated by choosing a specific spin configuration. The full spin dependence can be reconstructed by applying the relations in Eq.~\eqref{Eq::spin_elimination}. For the sake of simplicity we choose $(\sigma_i\sigma_k\sigma_j\sigma_l) = (\uparrow \downarrow \downarrow \uparrow)$ and redefine $P_{ij}^{b_ib_j} = V^P(i,i+b_i;j,j+b_j)$ and analogously for $C$ and $D$. The diagrammatic flow equations now read 
\begin{center}

    {
\scalebox{.7}
{

\begin{tikzpicture}
\node[] at(-1,0.5){$\frac{d}{d\Lambda}$};
\draw [fill = gray] (0,0) rectangle (0.25,1);
\draw (-0.5,0)  --node {\midarrow}(0,0) ;
\draw (-0.5,1)  --node {\midarrow}(0,1) ;
\draw (0.25,0)  --node {\midarrow}(0.75,0);
\draw (0.25,1) --node {\midarrow}(0.75,1);

\node[] at(1.25,0.5){$=$};

\node[] at(3,0.2){\Pch};

\node[] at(5,0.5){$+$};

\node[] at(7,0.2){\Cch};

\node[] at(1.25,-1.9){$+$};

\node[] at(5,-2.4){\Dch};

\end{tikzpicture}}
}

\end{center}

To write out the equations explicitly, we use Kronecker-deltas spanning the lattice around a specific site as a basis, defined as $f_{b_i}(k) = \delta_{i+b_i,k}$. We additionally assume a density-density type interaction. The sums in {Eq.~(\ref{Eq::Pproj}, \ref{Eq::Cproj}, \ref{Eq::Dproj})} can be reinterpreted as matrix-products in the multi-indices consisting of orbital and bond index. Thus, the flow equations now simplify to
\begin{align}
\begin{split}
    \frac{dP}{d\Lambda}  &= -\hat{P}[\Gamma] \cdot \dot{\chi}_{pp} \cdot \hat{P}[\Gamma] \\
    \frac{dC}{d\Lambda} &= -\hat{C}[\Gamma] \cdot \dot{\chi}_{ph} \cdot \hat{C}[\Gamma] \\
    \frac{dD}{d\Lambda} &= 2\hat{D}[\Gamma] \cdot \dot{\chi}_{ph} \cdot \hat{D}[\Gamma]\\ &- \hat{C}[\Gamma] \cdot \dot{\chi}_{ph} \cdot \hat{D}[\Gamma] \\ &-\hat{D}[\Gamma] \cdot \dot{\chi}_{ph} \cdot \hat{C}[\Gamma]
\end{split}
\end{align}
where $\dot{\chi}_{pp}$ and $\dot{\chi}_{ph}$ are redefined in terms of the spin summation. The Matsubara sum can be calculated either analytically which scales like $\mathcal{O}(N^4 N_b^4)$(with $N$ the number of orbitals) or numerically as a summation over the fermionic frequencies which scales proportional to $\mathcal{O}(N^2 N_b^2 N_f)$ (with $N_f$ the number of frequencies and $N_b$ the average number of bonds included) with an additional factor of $\mathcal{O}(N^3 N_f)$ for the calculation of the Greensfunction, which can be cached and therefore are  only calculated once. The number of frequencies can be reduced by employing a non-uniform Matsubara grid to approximately $10^2-10^3$ for reaching an accuracy below $10^{-5}$ at $T=10^{-3}$, rendering the numerical approach faster than the semi-analytical. 
The propagators are given as
\begin{align}
    \dot{\chi}_{ph  (i,j)}^{b_i,b_j} &= 2\sum_{\omega>0} \Re{\left(G(\omega)_{i,j}S(\omega)_{j+b_j,i+b_i} + G \leftrightarrow S\right)}, \\
    \dot{\chi}_{pp  (i,j)}^{b_i,b_j} &= 2\sum_{\omega>0} \Re{\left(G(\omega)_{i,j}S(-\omega)_{i+b_i,j+b_j} + G \leftrightarrow S\right)}.
\end{align}
The equation for the $D$-channel can be simplified by a completion of the square in its flow equations resulting in (defining $V_D = \hat{D}[\Gamma]-\frac{1}{2}\hat{C}[\Gamma]$)
\begin{equation}
    \frac{dD}{d\Lambda} = 2 V_D \dot{\chi}_{ph}V_D + \frac{1}{2}\frac{d}{d\Lambda} C
    \label{Eq::compsq},
\end{equation}
reducing the imbalance between the three channels greatly.
With the before defined Kronecker basis, the projections simplify due to the cancellation of sums. This results in Eq.~\eqref{Eq::Proj_1}-\eqref{Eq::Proj_3} (compare to \cite{weidinger_functional_2017, bauer_functional_2014}) introducing the difference set $\delta^d_{i,j}$ being one if the bond connecting $i$ and $j$ is included in the truncation.
\begin{align}
    \hat{P}[\Gamma]_{i,j}^{b_i,b_j} &= P_{i,j}^{b_i,b_j} +  \delta^d_{i,j+b_j}\delta^d_{i+b_i,j}C_{i,j}^{(j+b_j-i),(i+b_i-j)} +\delta^d_{i,j}\delta^d_{i+b_i,j+b_j}(D_{i,j+b_j}^{(j-i),(i+b_i-j-b_j)}+ U_{i,j+b_j}^{(j-i),(i+b_i-j-b_j)}) \label{Eq::Proj_1}\\
    \hat{C}[\Gamma]_{i,j}^{b_i,b_j} &= C_{i,j}^{b_i,b_j} +  \delta^d_{i,j+b_j}\delta^d_{i+b_i,j}P_{i,j}^{(j+b_j-i),(i+b_i-j)} +\delta^d_{i,j}\delta^d_{i+b_i,j+b_j}(D_{i,i+b_i}^{(j-i),(j+b_j-i-b_i)}+ U_{i,i+b_i}^{(j-i),(j+b_j-i-b_i)}) \label{Eq::Proj_2} \\
    \hat{D}[\Gamma]_{i,j}^{b_i,b_j} &= D_{i,j}^{b_i,b_j}  +  \delta^d_{i,j+b_j}\delta^d_{i+b_i,j}P_{i,i+b_i}^{(j+b_j-i),(j-i-b_i)} +\delta^d_{i,j}\delta^d_{i+b_i,j+b_j}C_{i,i+b_i}^{(j-i),(j+b_j-i-b_i)}+ U_{i,j}^{b_i,b_j} \label{Eq::Proj_3}.
\end{align}

\section{Decoupling of Vertex Function}
\label{App::B}
At the final scale, we obtain an effective interaction for the low lying energy degrees of freedom, with which we can reformulate the Hamiltonian as:
\begin{equation}
H = H_0 -\frac{1}{4} \Gamma(\alpha,\beta;\gamma,\delta)\bar{\psi}_{\alpha} \bar{\psi}_{\beta} \psi_{\gamma} \psi_{\delta}
\end{equation}
where we already went back to the Grassmann notation of the fermionic operators and introduced the multi-indices $\alpha = (i,\sigma_i)$. A post-FRG mean-field theory can enable a differentiation between competing orders, which are not separated by the FRG. But as the two approaches are decoupled, the resulting gap magnitudes depend on the stopping scale and the results should be therefore only be seen as a qualitative ordering pattern.
We now want to derive mean-field equations for the charge and spin gap. We restrict the derivation to a general, $SU(2)$-symmetric vertex and rewrite the effective interaction in the channel decomposed form:
\begin{align*}
\Gamma(\alpha,\beta;\gamma,\delta) = U_{\alpha,\beta;\gamma,\delta} + \Phi^C(\alpha,\beta;\gamma,\delta) + \Phi^D(\alpha,\beta;\gamma,\delta)
\end{align*}

where we neglected the pairing channel as for the cases in which we applied the mean-field decoupling it was suppressed with respect to the other channels. Additionally, we neglect the bare interaction, as in a flow to strong coupling its influence is small and therefore negligible.
In order to sum up the spin indices we again apply the relations in Eq.~\eqref{Eq::spin_elimination}. $C \text{ and } D$ are the channels we obtain as a result of our FRG scheme.
We start with the Grassmann path integral for the partition function:
\begin{align}
\mathcal{Z} = \int \mathcal{D}[\bar{\psi},\psi] &e^{S_0[\bar{\psi}\psi]} \cdot e^{-\frac{1}{4}\bar{\psi}_{\alpha}\bar{\psi}_{\beta} \Phi^C(\alpha,\beta;\gamma,\delta)\psi_{\gamma}\psi_{\delta}} \cdot e^{-\frac{1}{4}\bar{\psi}_{\alpha}\bar{\psi}_{\beta} \Phi^D(\alpha,\beta;\gamma,\delta)\psi_{\gamma}\psi_{\delta}}
\end{align}
We now introduce a matrix notation for the vertex components without spin degrees of freedom:
\begin{align*}
I(j,k;i,l) = I_{i,j}^{l,k}
\end{align*}
note that here we still have four spatial indices and no truncation has been applied yet.
We will now write out the spin dependence explicitly using the relations from Eq.~\eqref{Eq::spin_elimination}
\begin{align*}
\mathcal{Z} = \int \mathcal{D}[\bar{\psi},\psi] e^{S_0[\bar{\psi},\psi]} \cdot  & \exp \left[-\frac{1}{4} \bar{\psi}_j^{\sigma'} \bar{\psi}_{k}^{\sigma}  C_{i,j}^{l,k} \psi_{l}^{\sigma'} \psi_i^{\sigma}  + \frac{1}{4}\bar{\psi}_j^{\sigma}\bar{\psi}_{k}^{\sigma'} D_{l,j}^{i,k} \psi_{l}^{\sigma'} \psi_i^{\sigma} \right] \\
\cdot  &\exp \left[-\frac{1}{4} \bar{\psi}_{j}^{\sigma'} \bar{\psi}_{k}^{\sigma} D_{i,j}^{l,k} \psi_{l}^{\sigma'} \psi_i^{\sigma}  +\frac{1}{4} \bar{\psi}_j^{\sigma}\bar{\psi}_{k}^{\sigma'} C_{l,j}^{i,k} \psi_{l}^{\sigma'} \psi_i^{\sigma}\right].
\end{align*}
In order to have one type of index ordering per channel we commute the Grassmann variables and relabel the indices. The commutation of the fields results in an additional minus sign. We define the fermionic bilinears for each channel as:
\begin{align*}
C \implies \chi_{i,j}^{\sigma ,\sigma'} &=\; \bar{\psi}_i^{\sigma}  \psi_j^{\sigma'} \\
D \implies \rho_{i,j}^{\sigma, \sigma'} &=\; \bar{\psi}_i^{\sigma} \psi_j^{\sigma'} \delta_{\sigma \sigma'}
\end{align*}
Note that we have to use the physical channels in order to correctly assign all contributions to the respective mean-field. Those physical channels are defined as \cite{husemann_efficient_2009}
\begin{align*}
   M_{i,j}^{l,k} = \Phi^M(j,k;i,l) &= -\Phi^C(j,k;i,l) \\
   K_{i,j}^{l,k} =  \Phi^K(j,k;i,l) &= 
    2\Phi^D(j,k;i,l)-\Phi^C(j,k;i,l).
\end{align*}
$\Phi^M$ defines the magnetic-channel, a divergence of it thus describes transitions to a magnetic phase. $\Phi^K$ describes the charge-channel indicating charge ordering. 
The charge-channel does not contain a spin divergence part anymore, unlike the $D$-channel. Thus if both channels, $C$ and $D$, diverge the charge-channel does only contain the charge divergence which leads to an unambiguous decoupling.

Thereby, the full $SU(2)$-symmetric effective interaction is given as
\begin{align*}
\Gamma_{i,j}^{l,k} = -M_{i,j}^{l,k} +\frac{1}{2}(K_{i,j}^{l,k}-M_{i,j}^{l,k}).
\end{align*}
Upon inserting and reordering of the terms we obtain:
\begin{align*}
\mathcal{Z} = \int \mathcal{D}[\bar{\psi},\psi] e^{S_0[\bar{\psi},\psi]} 
\cdot  &  \exp \left[\frac{3}{4}M_{i,j}^{l,k}   \bar{\chi}_{i,j}^{\sigma,\sigma'}\chi_{k,l}^{\sigma,\sigma'}   \right] \\
\cdot  & \exp \left[-\frac{1}{4}  K_{i,j}^{l,k} \bar{\rho}_{l,j}^{\sigma'}\rho_{k,i}^{\sigma}   \right] 
\end{align*}
For the decoupling we now introduce two bosonic fields using a Hubbard-Stratonovitch transformation \cite{metzner_functional_2012}. In general this transformation reads (neglecting all occurring constants as we are only interested in the saddle point):
\begin{equation}
\exp\left[-\frac{1}{a} \bar{\eta}_{m} I_{m,n} \eta_{n} \right] = \int \mathcal{D}[\delta,\bar{\delta}] \exp\left[  \frac{1}{q}\bar{\delta}_mI_{m,n}\delta_n - \frac{b}{q}\bar{\delta}_m I_{m,n}\eta_n - \frac{b}{q}\delta_n I_{m,n}\bar{\eta}_m \right]
\end{equation}
With the restriction that $\frac{b^2}{q} = \frac{1}{a}$.
We label the bosonic fields $\phi$ with the subscript of their respective channel, $K$ for the charge-channel, $M$ for the magnetic-channel. Additionally to the bosonic fields, we introduce the energy gap as condensation term, defined by $\Delta^I = I\circ\phi^{I}$ ($\circ$ denotes a channel specific contraction of the tensor with the two fields, it will be specified later on, the charge gap contains an additional minus sign in its definition). 

\begin{align*}
\mathcal{Z} = \int& \mathcal{D}[\bar{\psi},\psi]  \mathcal{D}[\phi^M,\bar{\phi}^M]\mathcal{D}[\phi^K,\bar{\phi}^K]  \cdot \exp \left[S_0[\bar{\psi},\psi] + \frac{1}{2}\bar{\phi}^{M,\sigma,\sigma'}_{k,l}{M}_{i,j}^{l,k}\chi_{j,i}^{\sigma,\sigma'}  - \frac{1}{2}\bar{\phi}^{K,\sigma}_{j,l}K_{i,j}^{l,k} \rho_{k,i}^{\sigma'}  \right] \\
&\cdot\exp \left[ \frac{1}{2}\phi^{M,\sigma,\sigma'}_{k,l}{M}_{i,j}^{l,k}\bar{\chi}_{j,i}^{\sigma,\sigma'}  - \frac{1}{2}\phi^{K,\sigma}_{j,l} K_{i,j}^{l,k}\bar{\rho}_{k,i}^{\sigma'}   +\frac{1}{3} \bar{\Delta}^M_{j,i}\phi^M_{k,l} -\bar{\Delta}^K_{j,l}\phi^K_{k,i} \right] 
\end{align*}
By introducing the Hubbard-Stratonovitch field, the fermionic part of the action is reduced to a Gaussian and thus integrable. In the absence of magnetic fields we assume that $\phi^{M} = \phi^{M,\uparrow\uparrow} = -\phi^{M,\downarrow\downarrow}$, $\phi^{K} =\phi^{K,\uparrow\uparrow} = \phi^{K,\downarrow\downarrow}$ and $\phi^{M,\uparrow\downarrow} = \bar{\phi}^{M,\downarrow\uparrow}$. Additionally, the diagonal gaps must be self-adjoint. To simplify the notation we introduce the Nambu spinor
\begin{equation}
\mathbf{\Psi_i} = \begin{pmatrix}
   \psi^{\uparrow}_i& \psi^{\downarrow}_i
\end{pmatrix},
\end{equation}
with which we can rewrite the fermionic action in a compact form (note the reordering of Grassmann variables performed in order to have no minus signs occurring due to the integration):
\begin{align*}
\mathcal{Z}_{f} &= \int \mathcal{D}[\bar{\psi},\psi] \exp \left[\frac{1}{2}\mathbf{\Psi}_j (i\omega - \mathcal{M}_{ji})  \mathbf{\bar{\Psi}_i} \right]
\end{align*}
with the Matrix $\mathcal{M}$ defined as
\begin{align}
\mathcal{M} =\begin{pmatrix}
H + \Delta^{K}+ \Delta^{M}&  \Delta^{M,\uparrow,\downarrow} \\
\bar{\Delta}^{M,\uparrow,\downarrow} & H + \Delta^{K} - \Delta^{M} 
\end{pmatrix}.
\end{align}
Here we assumed that the magnetic gap will break the spin rotational invariance. The gaps will be initialized as a superposition of the leading eigenvectors of their respective relevant channel.

We now carry out the functional integrals, a Gaussian Grassmann integral, which results in (ignoring the prefactor as it has no effect on the saddle point equation)
\begin{align}
\mathcal{Z}_{f} &= \det\left(i\omega-\mathcal{M}\right) = \exp\left[\Tr\left(\ln\left(i\omega - \mathcal{M}\right)\right)\right] =  \exp\left[\Tr(\ln(G^{-1})) \right]
\end{align}
where we identified the Greens-function $G$. The mean-field equations are obtained by a saddle point approximation, defined as:
\begin{align}
\frac{\delta\mathcal{Z}[\phi]}{\delta\phi} = 0 \iff \frac{\delta S[\phi]}{\delta\phi} = 0
\end{align}
The variation of each field needs to vanish individually, leading to a coupled set of self-consistent equations. We sketch the derivation for the charge gap in the following (we suppress spin indices for brevity).
\begin{align}
0 &= \frac{\delta S[\phi^K,\bar{\phi}^K,\phi^{M,\sigma,\sigma'},\bar{\phi}^{M,\sigma,\sigma'}]}{\delta\phi^K_{ij}} \\ 
&= \Tr \left( G_{mn}\cdot \frac{\delta G^{-1}[\phi^K,\bar{\phi}^K,\phi^{M,\sigma,\sigma'},\bar{\phi}^{M,\sigma,\sigma'}]}{\delta\phi^K_{ij}}\right) - \left(\bar{\Delta}^K_{j,i} + \Delta^K_{j,i}  \right) \\ 
&= -2\Delta^{K}_{ij} - \sum_{\omega} \left(K_{n,j}^{i,m}(G(\omega)^{\uparrow,\uparrow}_{nm} + G(\omega)^{\downarrow,\downarrow}_{nm})\right)
\end{align}
The spin indices are marking the spin block indices of the Greens-function. This equation can now be solved for the charge gap:
\begin{equation}
\Delta^{K}_{lj} = -\frac{1}{2}K_{i,j}^{l,k} \sum_{\omega} (G_{ik}^{\uparrow,\uparrow}(\omega)+G_{ik}^{\downarrow,\downarrow}(\omega))
\end{equation}

The equation for $\Delta^{M}$ follows analogously
\begin{equation}
\Delta^{M,\sigma,\sigma'}_{ij} = \frac{3}{2}M_{i,j}^{l,k} \sum_{\omega} (G_{lk}^{\sigma,\sigma'}(\omega) +(-1)^{\delta_{\sigma,\sigma'}} G_{lk}^{-\sigma,-\sigma'}(\omega))
\end{equation}

\begin{SCfigure}
    \centering
    \includegraphics[width = 0.5\textwidth]{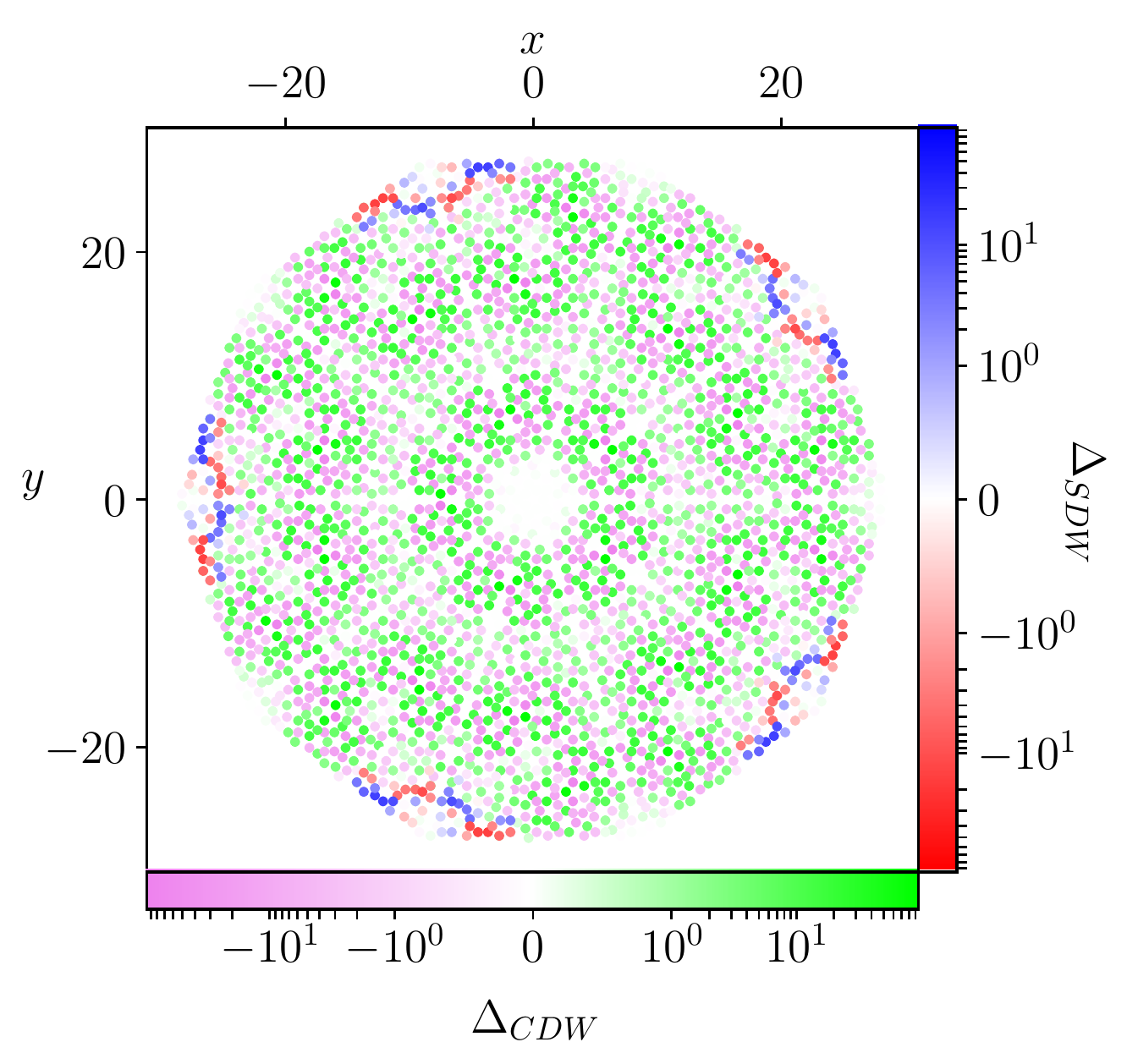}
    \caption{Gap magnitudes of SDW and CDW in the center Penrose model at $\mu=2.0$, $U=1.0$, $U' = 1.015$ and $T=10^{-3}$ without fixed particle number. It compares well to the gap predicted by the fixed particle number calculation. he calculated gap magnitudes are shown in a red-blue color scheme for the spin gap, which is equivalent to the magnetisation. In the violet-lime color scheme the charge gap is shown. The two faces have no spatial overlap of magnitudes above $10^{-3}{\rm max}(\Delta)$.}
    \label{fig:centergapnofix}
\end{SCfigure}

The channels we obtain from the TU calculation have the following index structure:
\begin{align}
C(i,k;j,l) &\rightarrow C(i,j+b_j;j,i+b_i)  \\
D(i,k;l,j) & \rightarrow D(i,j+b_j;i+b_i,j)  
\end{align}
which leads to the following form of the physical channels
\begin{align}
   M_{i,k}^{b_i,b_k} &= -\frac{3}{2}C_{i,k}^{b_i,b_k} \\
   K_{i,k}^{b_i,b_k} &= 
    2D_{i,k}^{b_i,b_k}-C_{i,k}^{b_i,b_k}.
\end{align}
We now return to the real-space TUFRG notation we introduced earlier, thus the index ordering is changed.
Using the eigenvector-decomposition which we obtain as result of our FRG flow we reconstruct the approximate vertex as:
\begin{align}
K(j,i+b_i;j+b_j,i) &= K_{i,j}^{b_i,b_j} = \sum_k \lambda_k \ket{k}_i^{b_i} \bra{k}_j^{b_j} \\
M(i+b_i,j;j+b_j,i) &= C_{i,j}^{b_i,b_j} = -\sum_c \lambda_c \ket{c}_i^{b_i} \bra{c}_j^{b_j}
\end{align}
This sum will be truncated to a few of the largest eigenvalue/eigenvector combinations indicated by a tilde from now on.
Here we still need to sum out the inner spin index as well as the Matsubara sum. As the vertex is frequency independent we can solve the Matsubara sum by complex integration
\begin{equation}
\sum_{\omega}G_{kl} = \sum_n U_{kn}n_f(\lambda_n) U_{ln}^* = n_f(\mathcal{M})_{kl}.
\end{equation}
The sum over all leading eigenvectors gives the effective channels resulting in the following set of self-consistent equations
\begin{align}
\Delta^{M}_{i,i+b_i} &=  \frac{3}{2}\tilde{M}_{ij}^{b_i b_j}(n_f(\mathcal{M})^{\uparrow,\uparrow}-n_f(\mathcal{M})^{\downarrow,\downarrow})_{j,j+b_j} \\
\Delta^{M}_{i,i+b_i} &=  \frac{3}{2}\tilde{M}_{ij}^{b_i b_j}(n_f(\mathcal{M})^{\uparrow,\downarrow}+n_f(\mathcal{M})^{\downarrow,\uparrow})_{j,j+b_j} \\
\Delta^{K}_{i,i+b_i} &=  -\frac{1}{2}\tilde{K}_{ij}^{b_ib_j}(n_f(\mathcal{M})^{\uparrow,\uparrow}+n_f(\mathcal{M})^{\downarrow,\downarrow})_{j,j+b_j}.
\end{align}
If only a single channel is diverging, we can reduce the number of mean-fields to obtain a more efficient description. In practice we keep the particle number constant during the self-consistency iteration by adding an appropriate, adaptive chemical potential. Additionally, we add and subtract the "zero gap charge gap" and absorb one part into this constant, the other part is used to redefine the charge gap
\begin{align}
\tilde{\Delta}^{K}_{i,i+b_i}=\Delta^{K}_{i,i+b_i}-\Delta^{K}_{i,i+b_i}|_{\Delta^{K}=0} &=  -\frac{1}{2}\tilde{K}_{ij}^{b_ib_j}(n_f(\mathcal{M})^{\uparrow,\uparrow}+n_f(\mathcal{M})^{\downarrow,\downarrow} -\delta_{0,b_j}( n_f(H)^{\uparrow,\uparrow}+n_f(H)^{\downarrow,\downarrow}))_{j,j+b_j},
\end{align}
in order to make its value vanish if no divergence is encountered in the physical charge-channel.
The fixing of the particle number introduces convergence issues which are addressed by a mixing parameter (see Eq.~\eqref{Eq:mixing}).
\begin{equation}
    \Delta^{n+1} = \alpha\Delta^{n} + (1-\alpha)\Delta^{n-1}.
\label{Eq:mixing}
\end{equation}
Additionally, the fixing of the particle number introduces indirect coupling of the two regions, as particles cannot simply be pushed out of the one region. They then flow into the other region creating a charge displacement there. We always performed both simulations, once for fixed particle number and once for non-fixed particle number to check convergence. 
The non coupled MF/FRG leads to very large gap magnitudes which introduce additional convergence issues. This is resolved by a rescaling of the vertices by a factor, we used $\nicefrac{1}{10}$. As the resulting gap magnitudes are only qualitative this is not introducing a bias. In Fig. \ref{fig:centergapnofix} we show the gap magnitudes for the non particle number fixed case of the coexistence of magnetic and charge ordering in the center Penrose model.

\section{Interaction envelop}
\label{App::C}

\begin{SCfigure}
    \centering
    \includegraphics[width =0.45\textwidth]{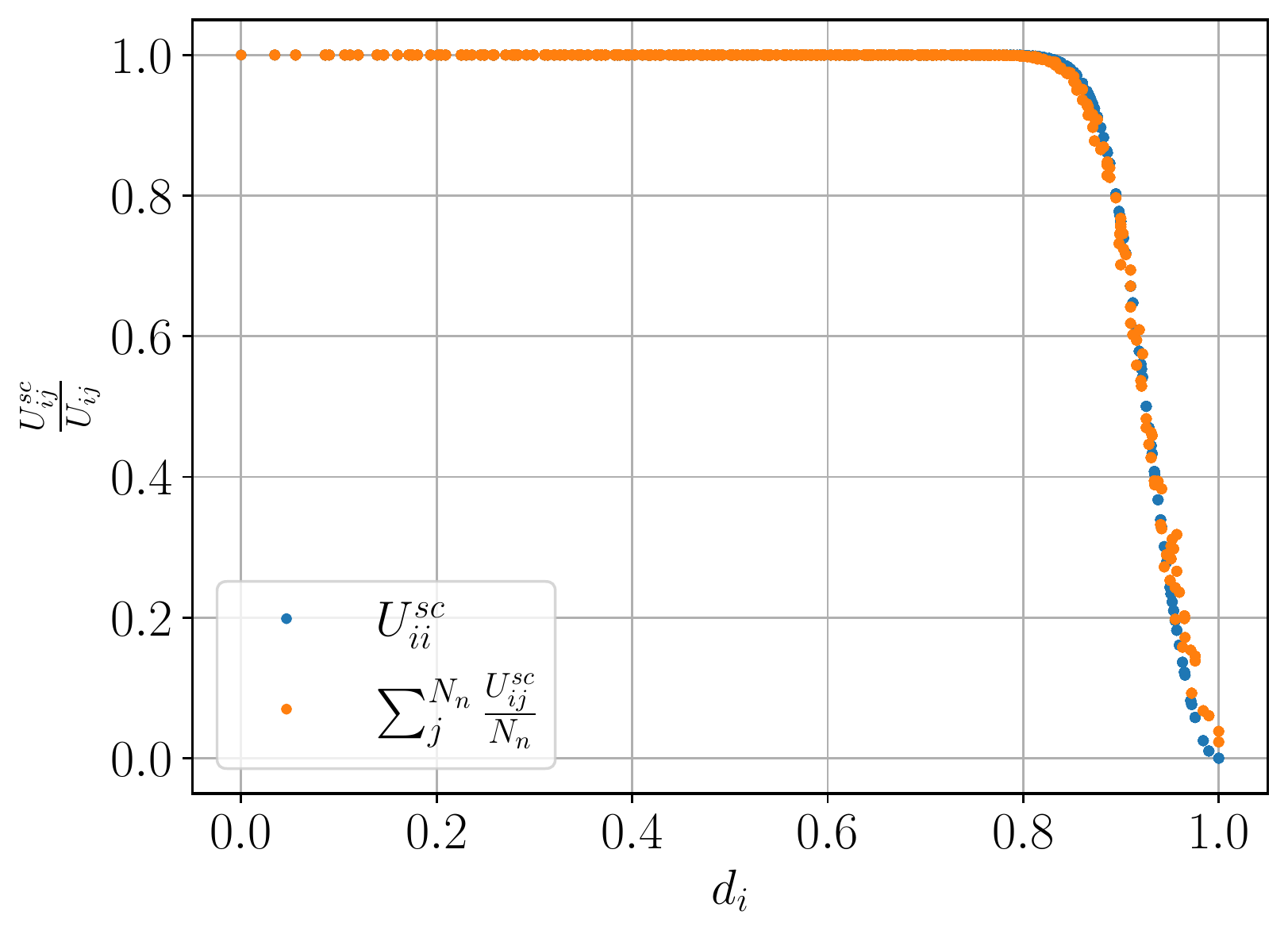}
    \caption{Average on-site and nearest-neighbor interaction depending on the distance defined by the real-space positions to the center in the vertex-type Penrose model.}
    \label{fig:distance_dep_screening}
\end{SCfigure}

For the interaction envelop we apply
\begin{align}
 U_{ii}\to  U_{ii}^{sc} = U_{ii} \tanh(10\cdot(d_{max}-d_i)^2)
\end{align}
as scaling factor for the interaction ($d_i$ is the distance of site i to the center of the tiling and $d_{max} = \max_i d_i$). For the nearest-neighbor interactions, this factor is applied too. To keep the interaction symmetric and $C^5$-invariant we need to symmetrize the scaling factor (see Eq.~\eqref{Eq:sym_scal}), thereby we introduce a slight bias as explained below.
\begin{align}
    U_{ij}^{sc} =\frac{U_{ij}}{2}\cdot\left[\tanh(10\cdot(d_{max}-d_i)^2)+\tanh(10\cdot(d_{max}-d_j)^2)\right]
\label{Eq:sym_scal}
\end{align}
This creates a region of higher $\nicefrac{U'}{U}$-ratio than in the bulk. This slight deviation seems not to have any influences as each interaction term is lower individually. The effective lattice size is of course reduced due to the application of the screening. The distance dependence of the screened on-site and nearest-neighbor interaction is shown in Fig.~\ref{fig:distance_dep_screening}.

\section{Non interacting Density of states}
\label{App::D}
The non-interacting Density of states (DoS) has been calculated for the 9-times substituted lattice. We used the Kernel Polinomial Approximation method. The DoS was used in order to choose the parameters for the simulations. We focus on regions with high degeneracy or low slope of the dispersion relation in the diagonal basis of the Hamiltonian, both lead to strong peaks in the DoS.
\begin{figure}[!htbp]
\centering
\includegraphics[width = 0.32\linewidth]{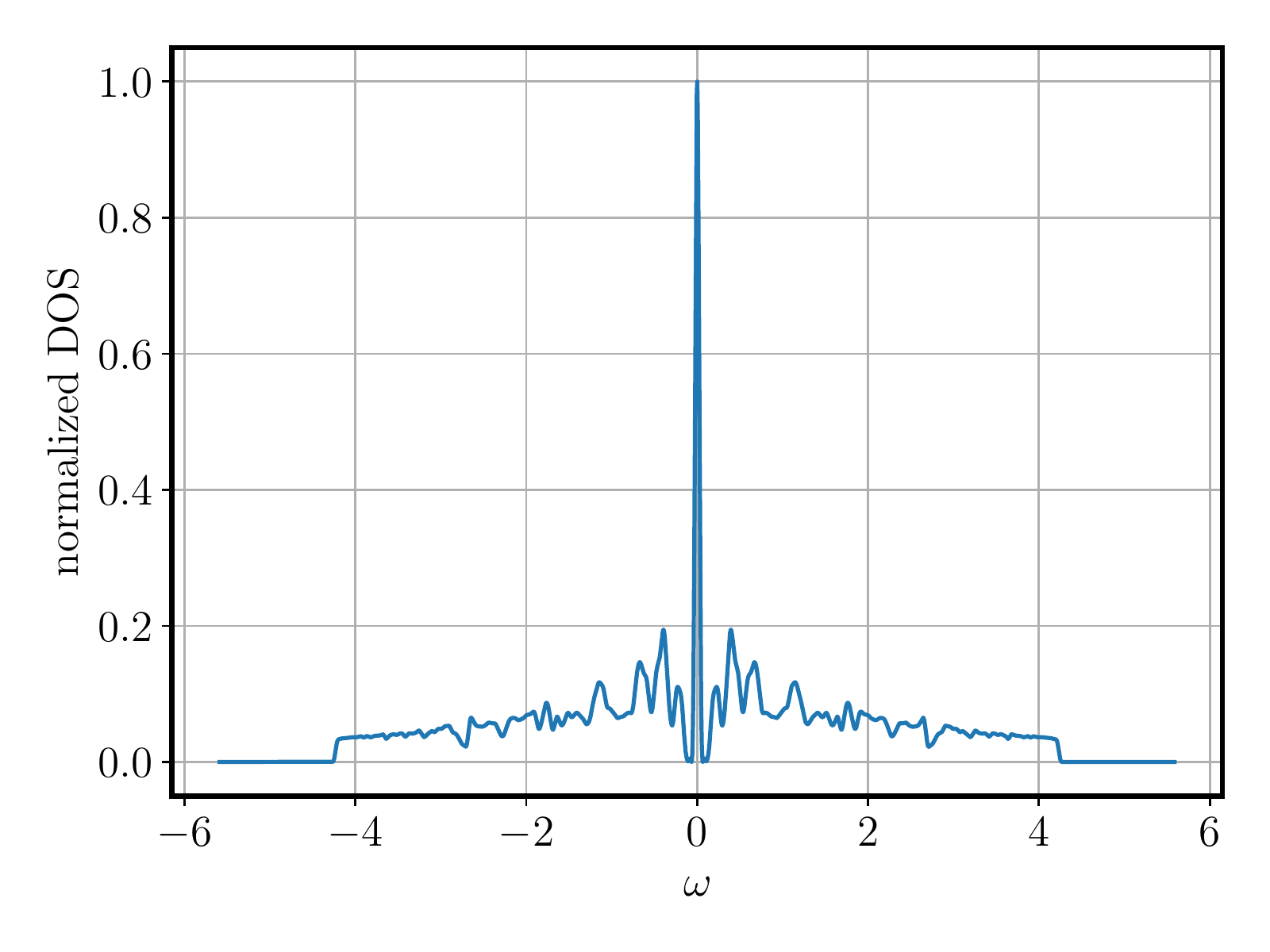}
\hfill
\includegraphics[width = 0.32\linewidth]{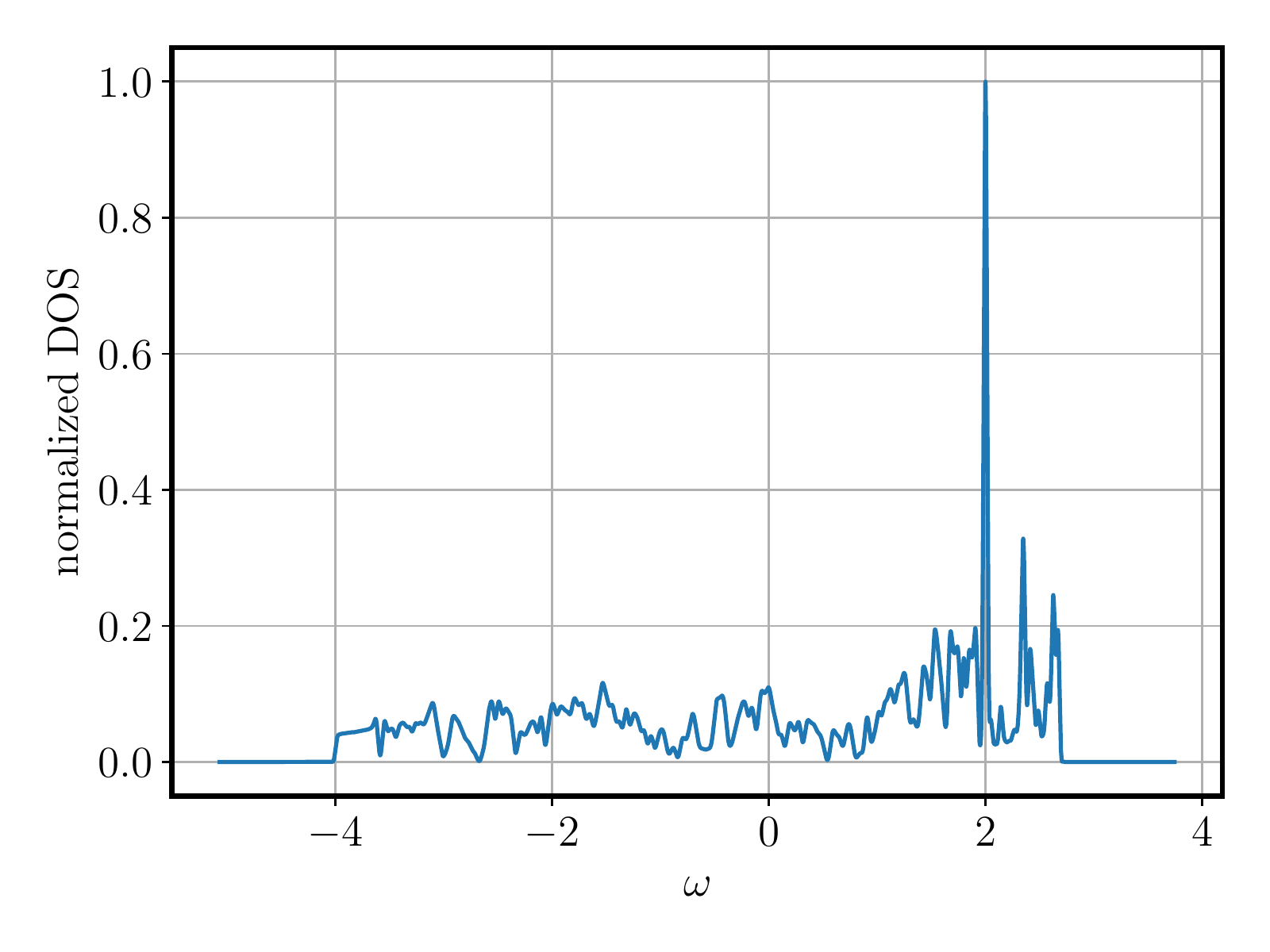}
\hfill
\includegraphics[width = 0.32\linewidth]{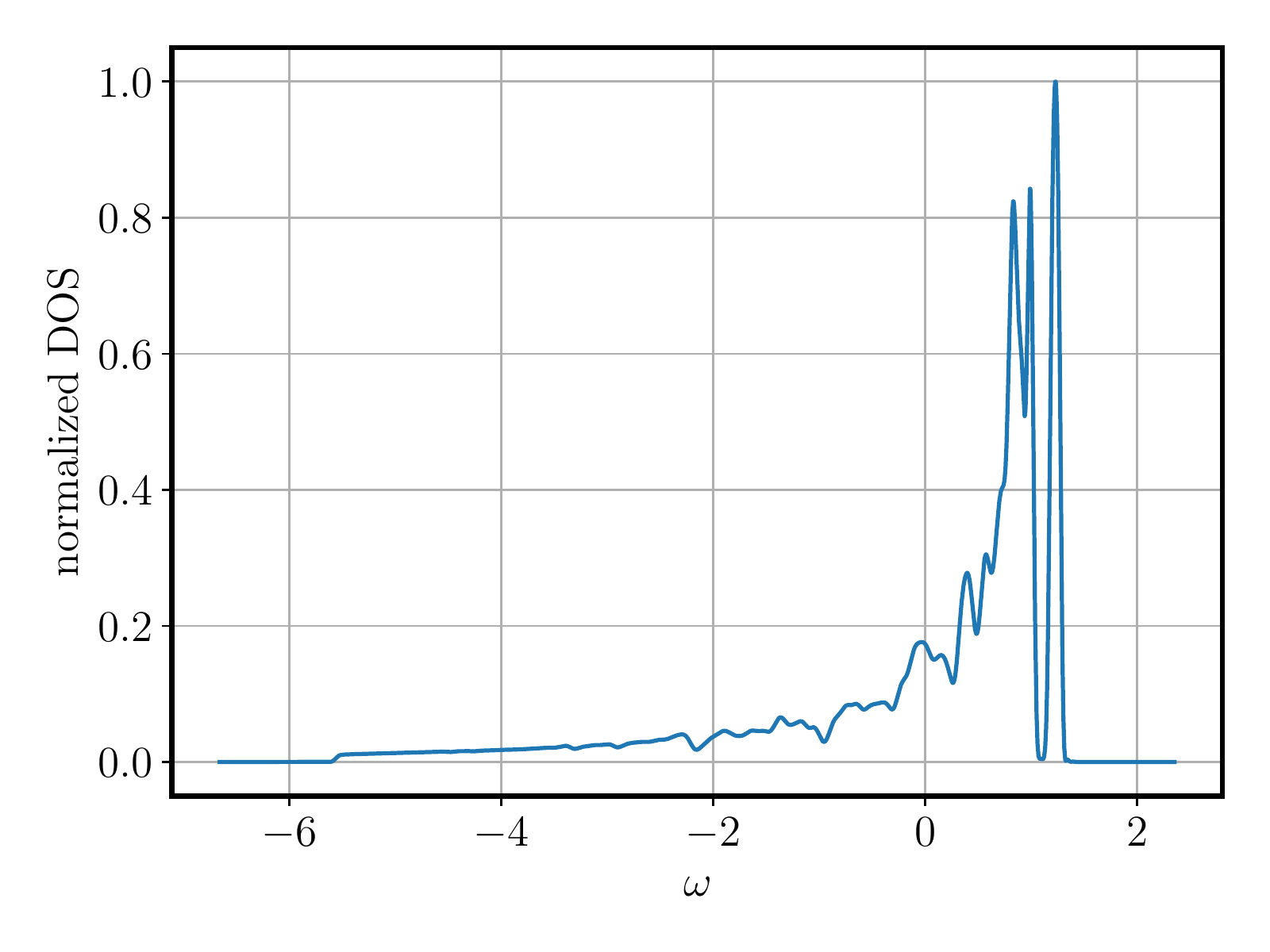}

\caption{Non interacting DoS for the three types of Penrose models used. Left, we have the vertex-type Penrose model with $21106$ sites with the $\delta$-like peak at $\mu=0$. In the middle, we show the center-type model with hoppings defined using the graph, the main peak is at $\mu = 2.0$. On the right, we show the center-type model with exponential decaying hoppings, with three main peaks. In both center type models the lattice contains $20800$ sites.}
\label{Fig::DOS}
\end{figure}

\section{Vertex model at half filling}
\label{App::E}
\begin{SCfigure}[][!htbp]
    \centering
    \includegraphics[width = 0.4\textwidth]{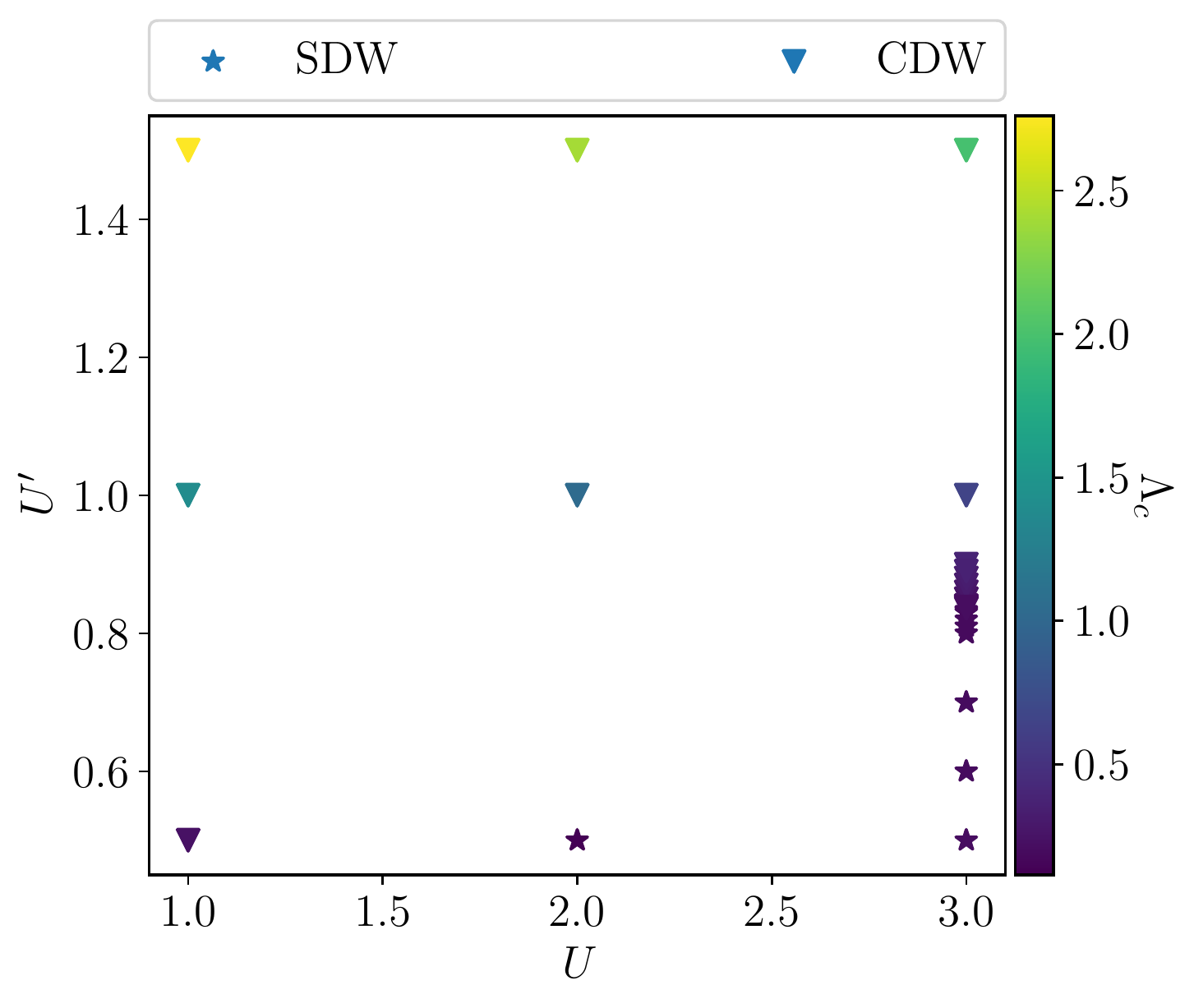}
    \caption{Phasediagram of the vertex Penrose model employing Eq.\eqref{Eq:sym_scal}. we find a SDW (star marker) and a CDW (triangular marker) phase. The critical scales are given by the color-code.}
    \label{fig:PD_vertex}
\end{SCfigure}

The phase diagram for the vertex model with employed envelop, discussed in the main boy is shown in Fig.~\ref{fig:PD_vertex}.

\begin{figure}[!htbp]
    \centering
    \includegraphics[width = 0.49\textwidth]{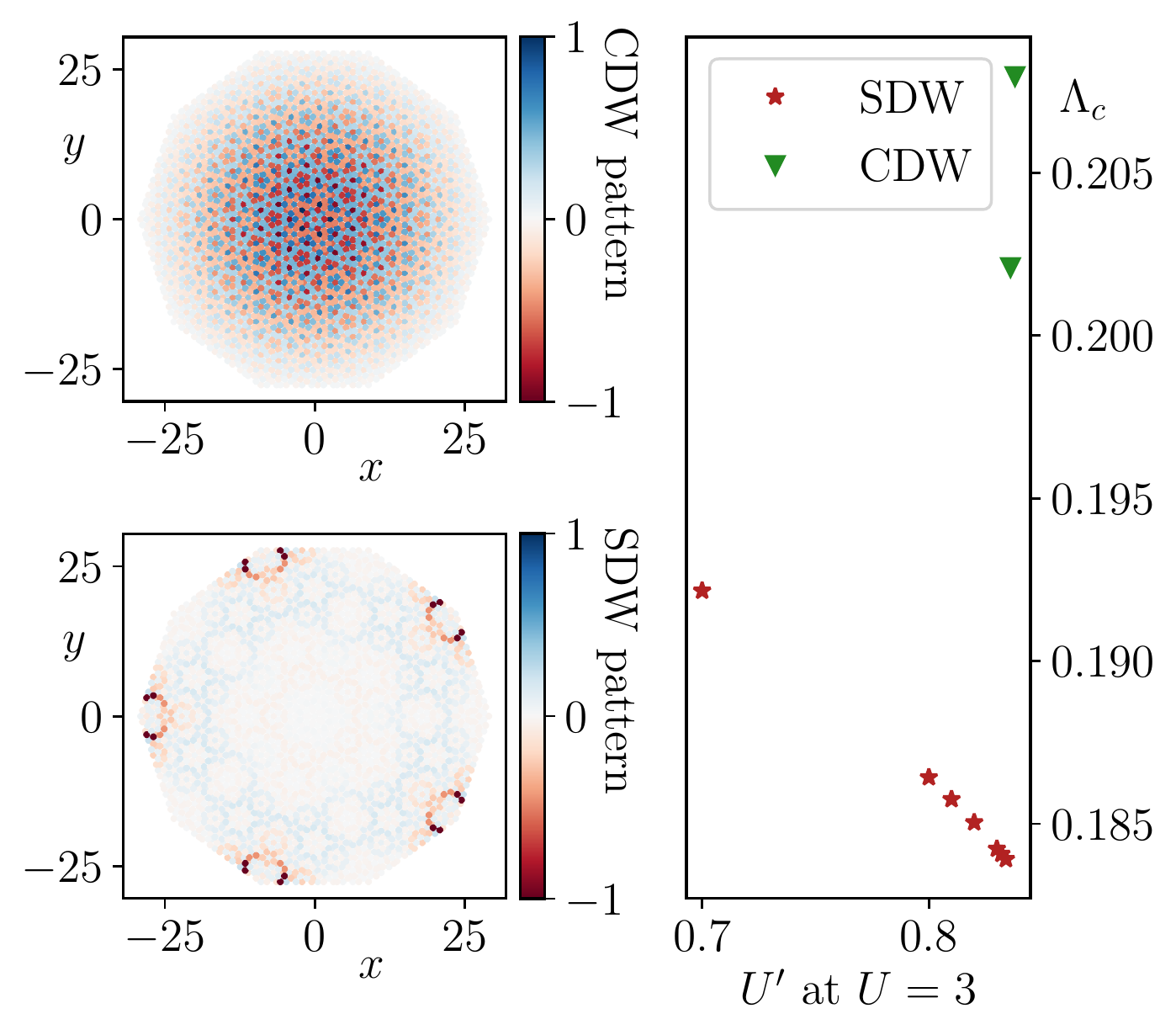}
    \includegraphics[width = 0.49\textwidth]{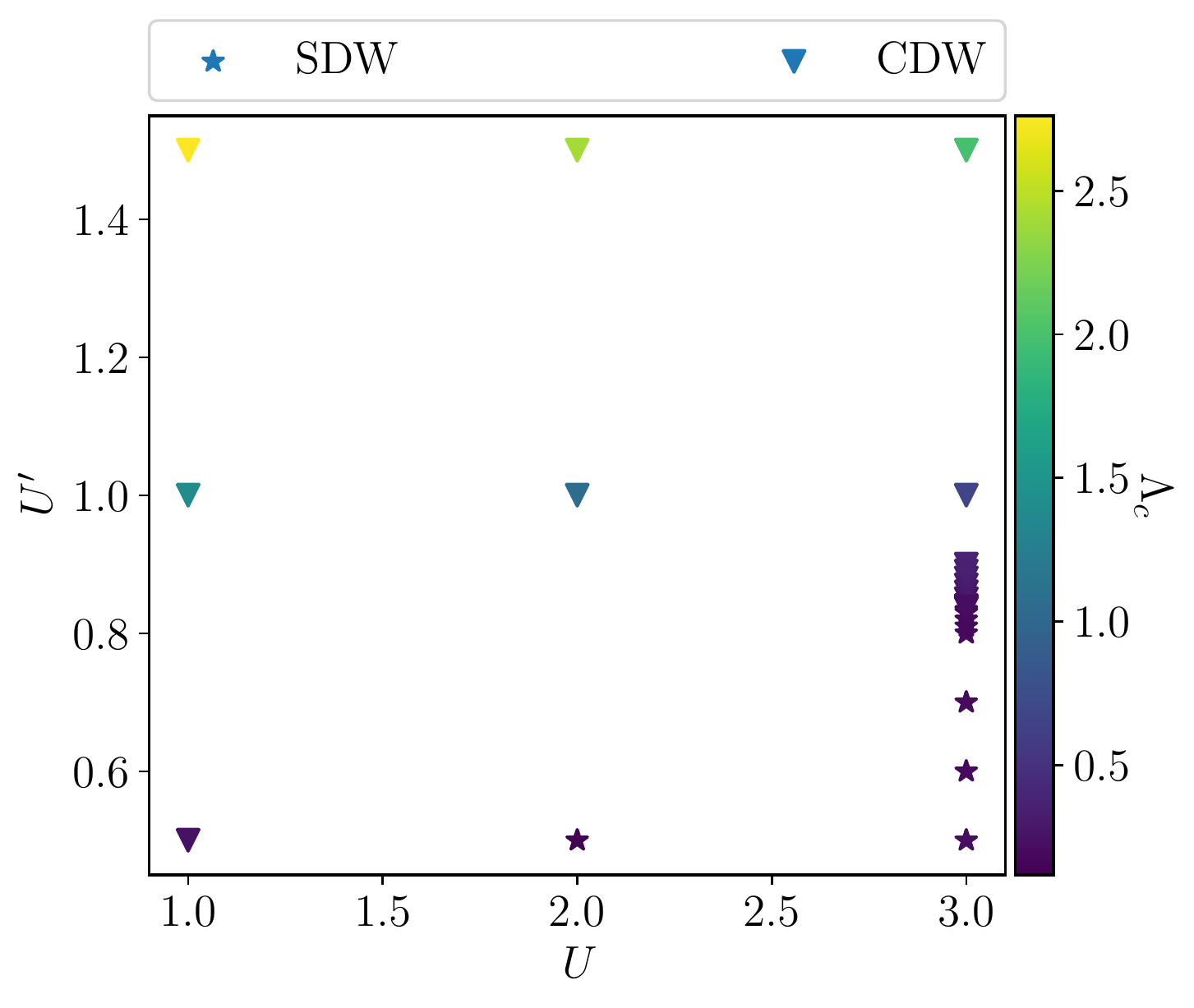}
    \caption{ The left plot shows in the upper left inset the charge order parameter at $U'=1.5$ and $U=1$. The lower left inset visualizes the spin order parameter (or magnetization pattern), at $U'=0.5$ and $U=2.0$. On the right side, the critical scales depending on the nearest-neighbor interaction are shown for $U = 3.0$ in the vicinity of the phase transition, to highlight the suppression of the critical scale upon approaching the phase transition. The right plot shows the phase diagram of the vertex model with a SDW (star marker) and a CDW (triangular marker) phase, the phase diagrams for the pure case and the one employing an envelop differ only very slightly.}
    \label{fig:no_screening_vertex}
\end{figure}

For pure interaction, the main results are summarized in Fig. \ref{fig:no_screening_vertex}.
We find a charge-density wave phase at dominant nearest-neighbor interactions ($U'$) and a spin-density wave phase at dominant on-site interactions. The latter is expected as in the limit of weak $U'$ we should recover the known antiferromagnet for the vertex model. The SDW phase has its main weight at the boundaries, which is a consequence of the open boundary conditions and can be understood as follows. At the boundaries the local kinetic energy bandwidth is reduced, which results in a larger $\nicefrac{U}{W_D}$ ratio, which in term leads to faster divergences of diagrammatic re-summations at the boundaries. The sub-leading eigenvectors of the SDW divergence also show this bulk behavior. Upon decreasing $U'$, this effect is also vanishing as it should be. The CDW phase loses weight towards the boundaries and thus is seen as a bulk phase. Due to the reduced boundary weight, it could be possible that a coexisting boundary phase emerges at lower critical scales or lower temperatures. However, we did not find a phase coexistence. At the transition between CDW and SDW we find a mutual suppression of the phases resulting from their non-zero overlap in space.

This behavior can be understood as follows. In a spin-channel (SDW) antiferromagnetic divergence the charge-channel (CDW) has an eigenvalue, describing the strength of the order, half as large as the one of the spin-channel, which already follows from the RPA. This eigenvalue is in our sign convention positive. The diverging eigenvalue of the CDW-RPA itself is negative, thus the two will suppress each other in the vertex reconstruction. This partial cancellation is the source of the reduced critical scales and thus reduced transition temperatures. As soon as the CDW divergence is dominant the critical scales grow rapidly again, as shown in the right inset of the left plot in Fig.~\ref{fig:no_screening_vertex}.

\section{Phase-diagram of the center model at different filings.}
\label{App::F}
\begin{figure}[!htbp]
\centering
{\includegraphics[width = 0.32\linewidth]{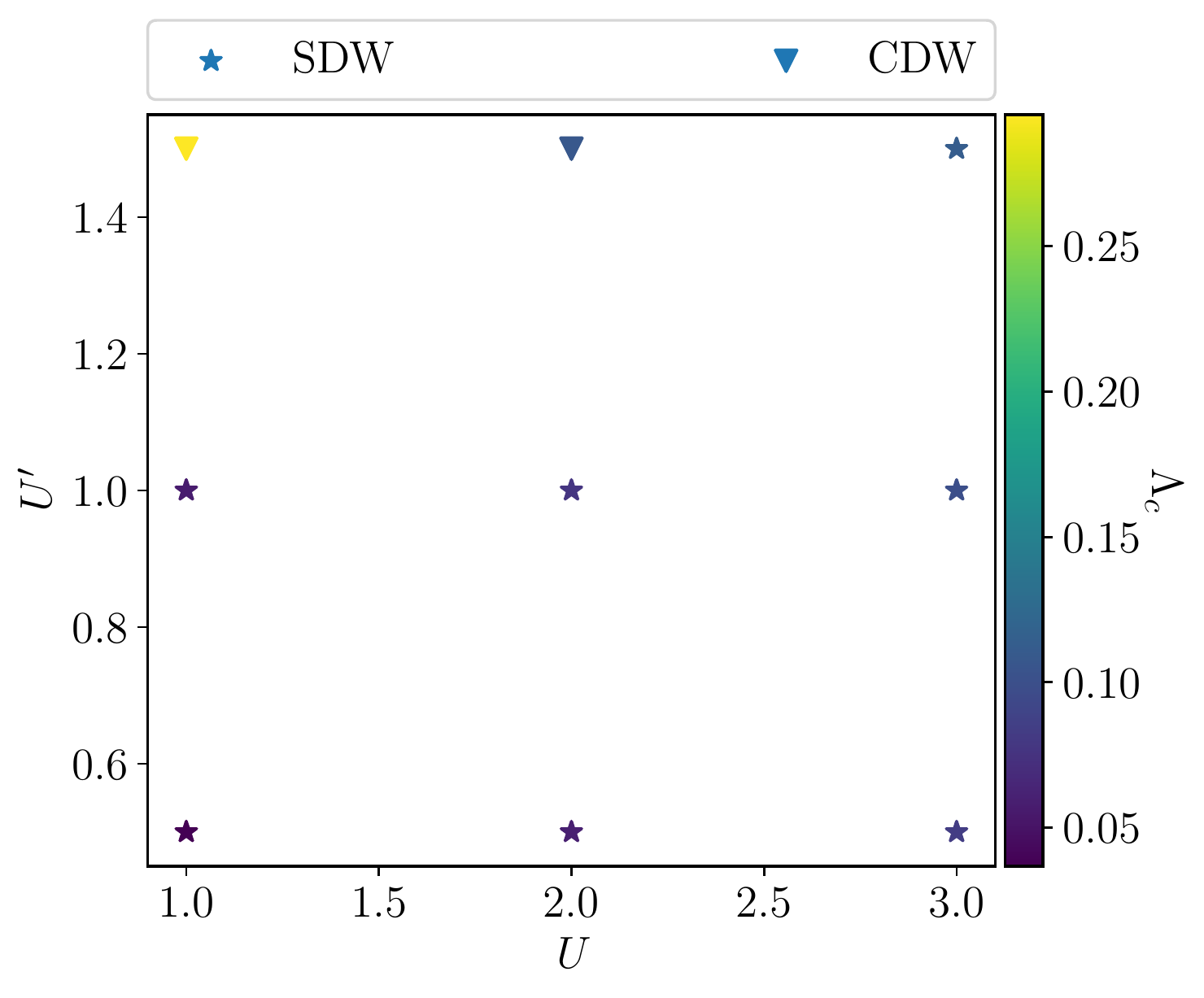}}
\hfill
{\includegraphics[width = 0.32\linewidth]{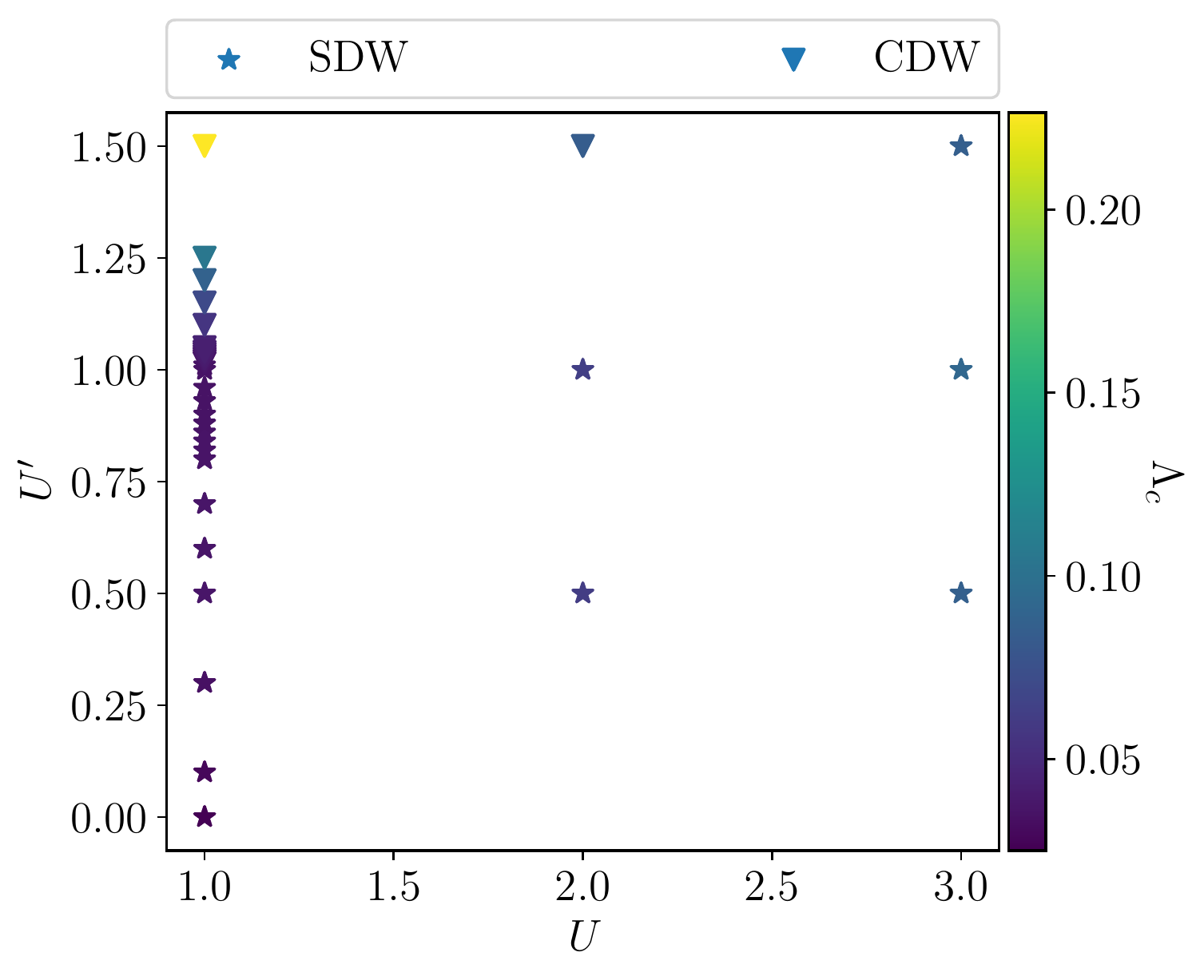}}
\hfill
{\includegraphics[width = 0.32\linewidth]{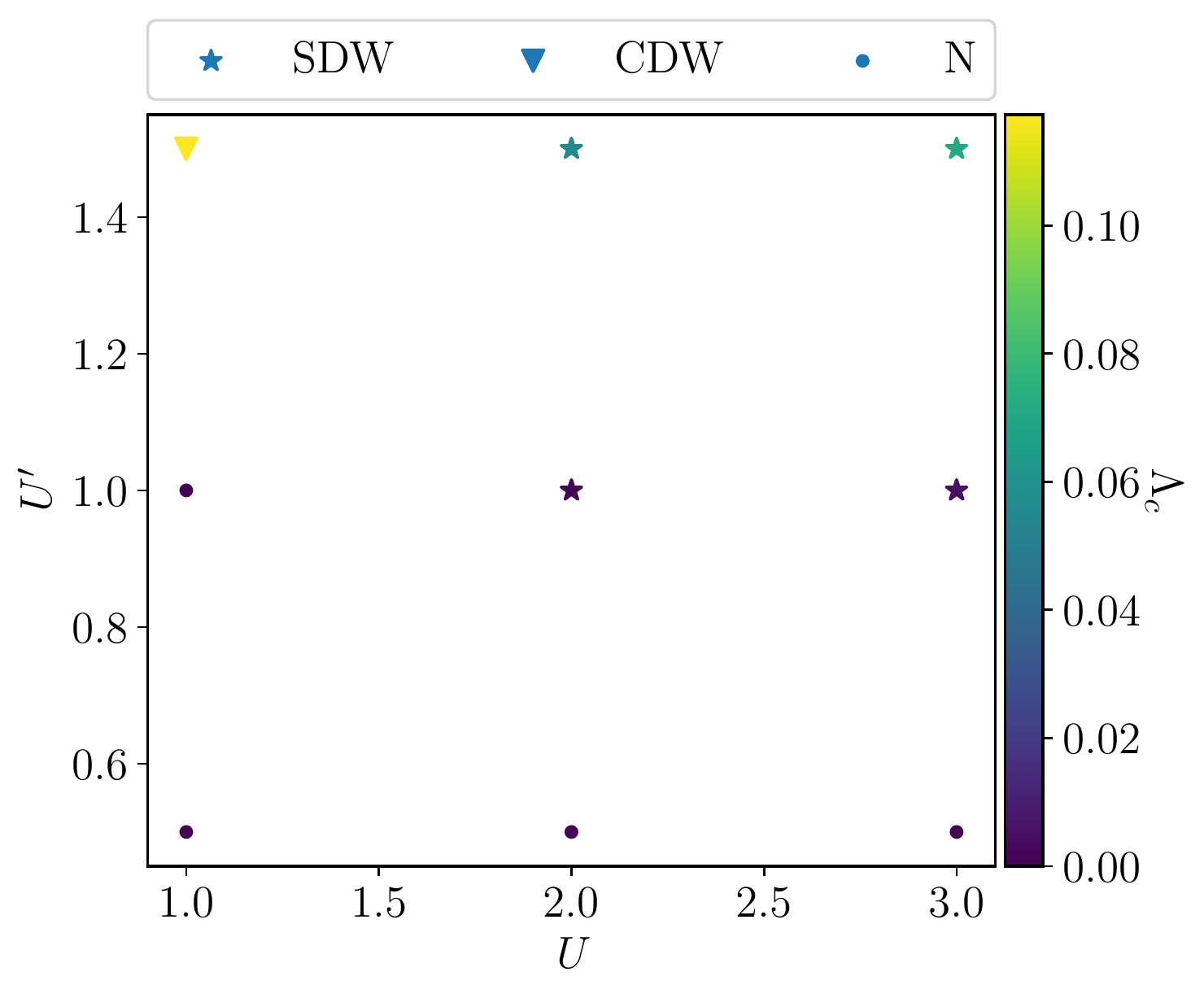}}
\vfill
{\includegraphics[width = 0.32\linewidth]{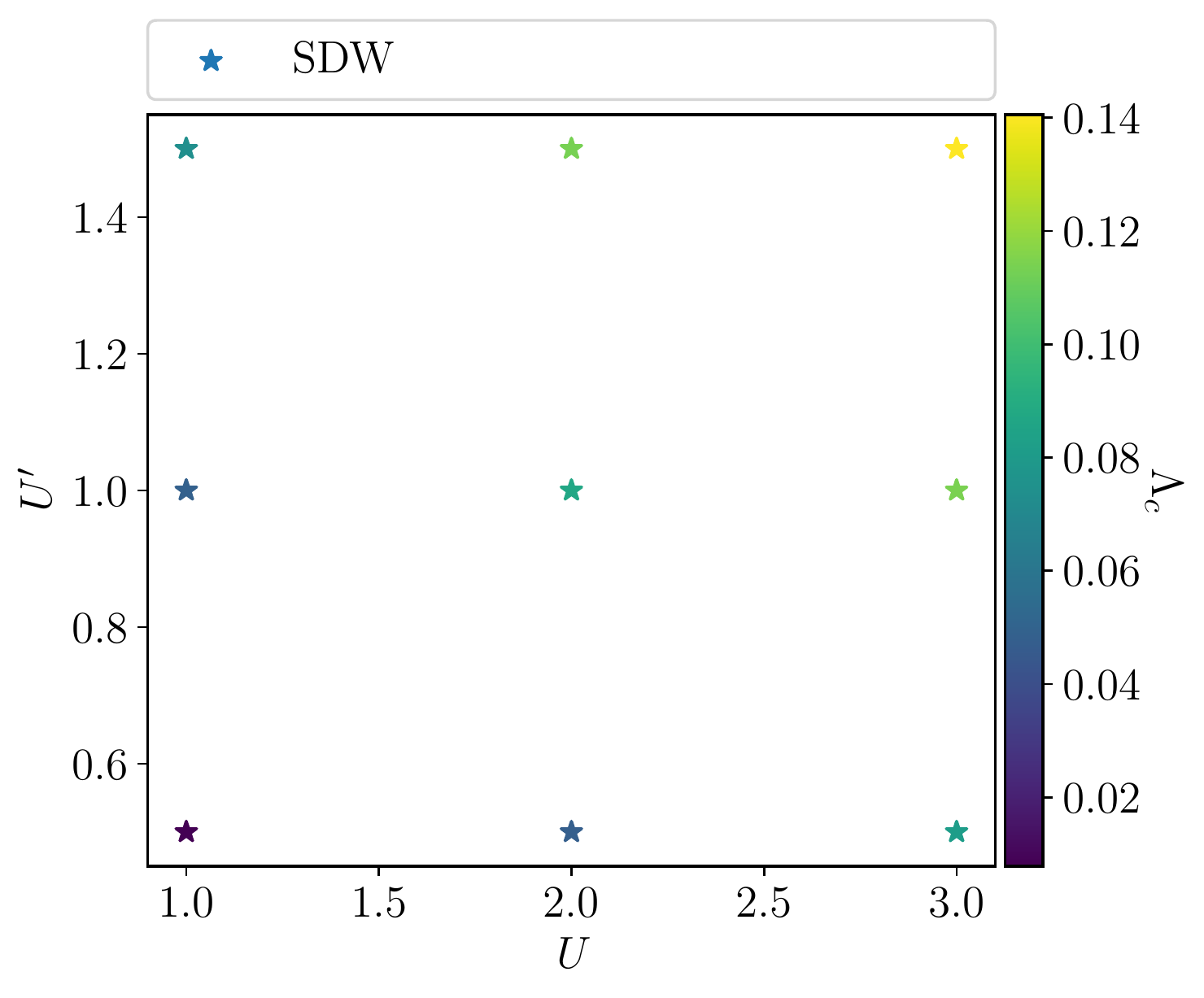}}
\hfill
{\includegraphics[width = 0.32\linewidth]{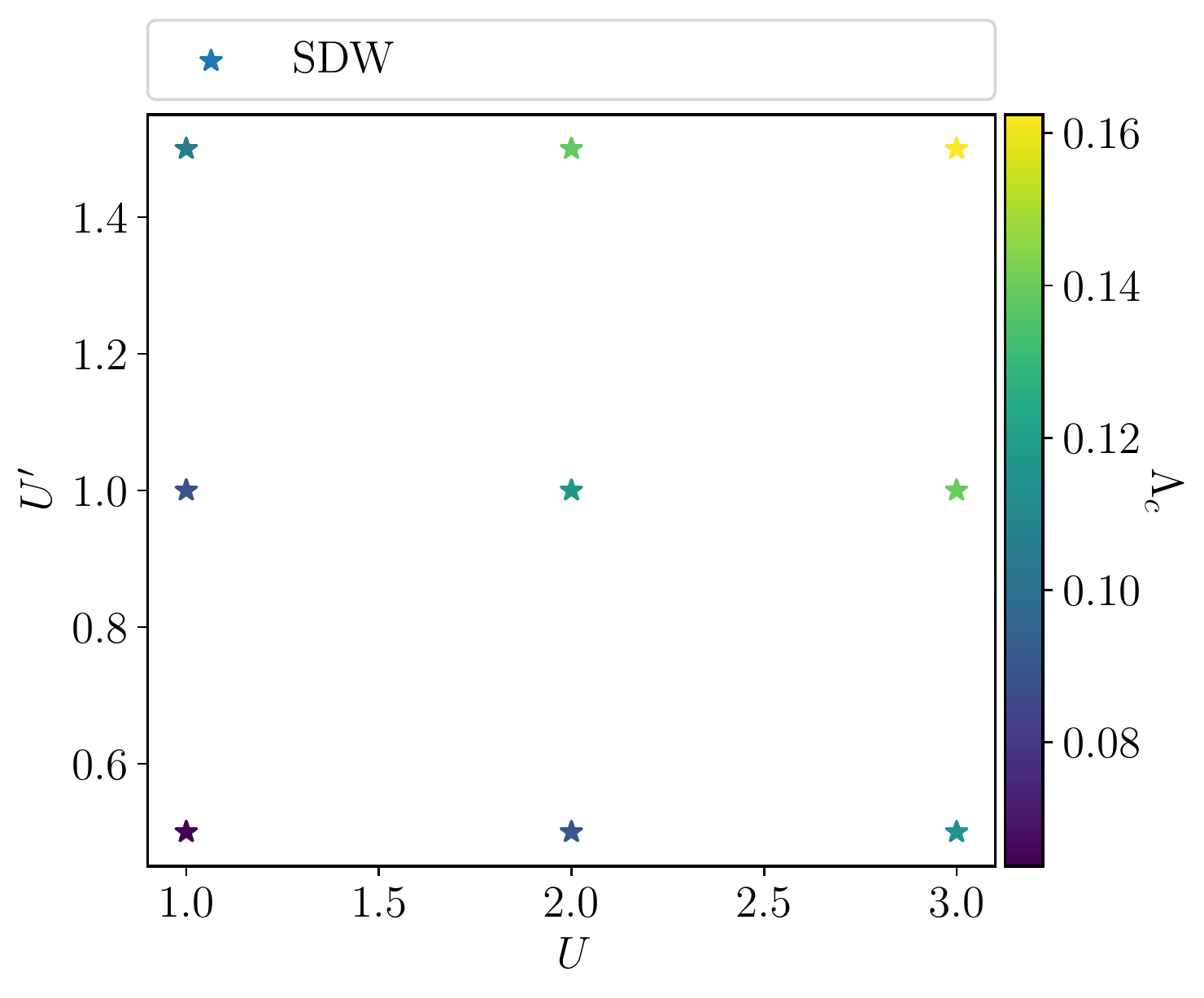}}
\hfill
{\includegraphics[width = 0.32\linewidth]{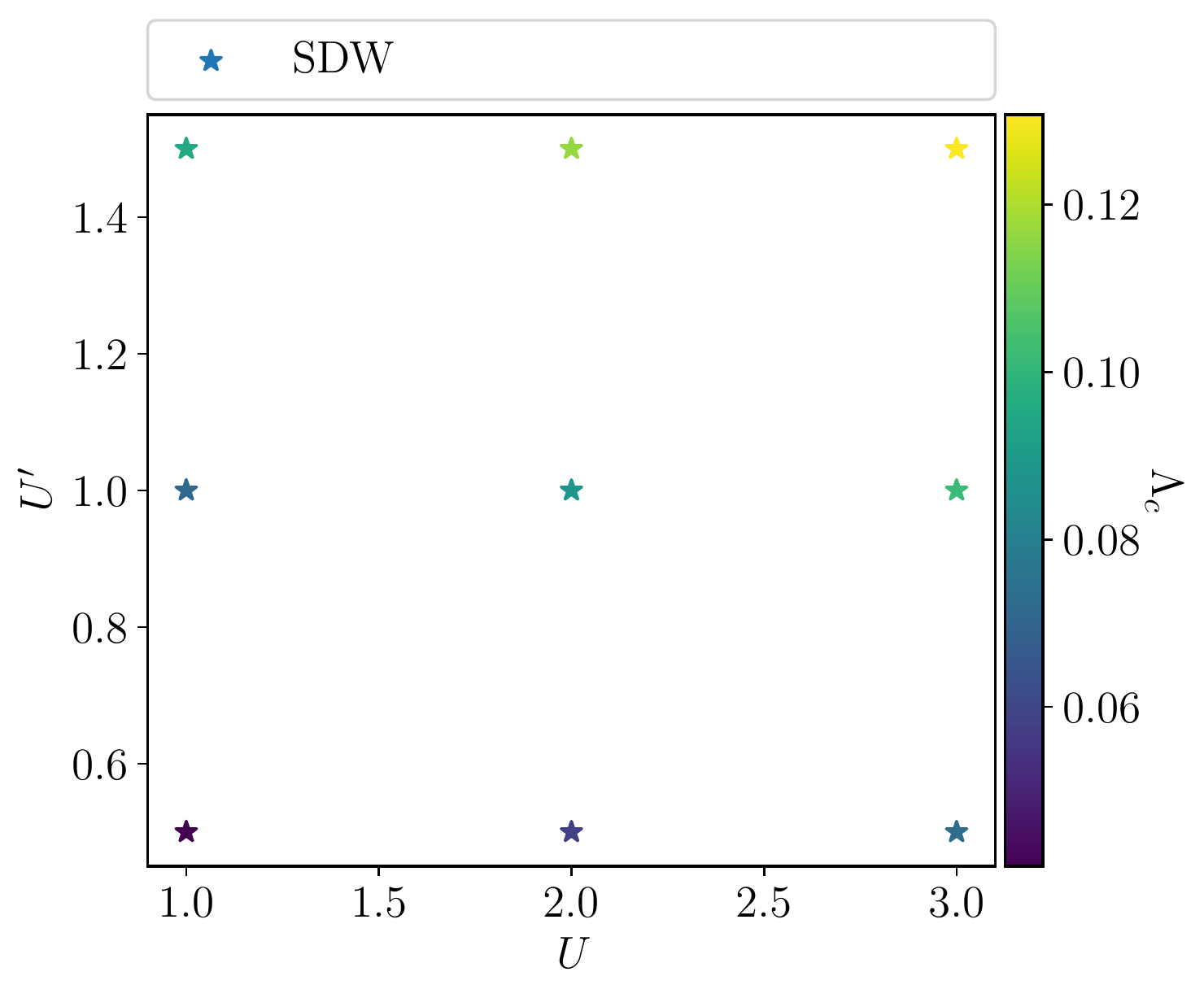}}

\caption{Phase-diagram of the center model at different chemical potentials, we have $\mu = 1.9$ in the upper left, $\mu = 2.0$ in the upper middle, $\mu = 2.1$ in the upper right, $\mu = 2.3$ in the lower left, $\mu = 2.35$ in the lower middle and $\mu = 2.4$ in the lower right plot. $U$ and  $U'$ are varied at $T=10^{-3}$, nearest neighbors are included in the calculation.}
\label{Fig::phases_center}
\end{figure}

\begin{figure}[!htbp]
\centering
{\includegraphics[width = 0.32\linewidth]{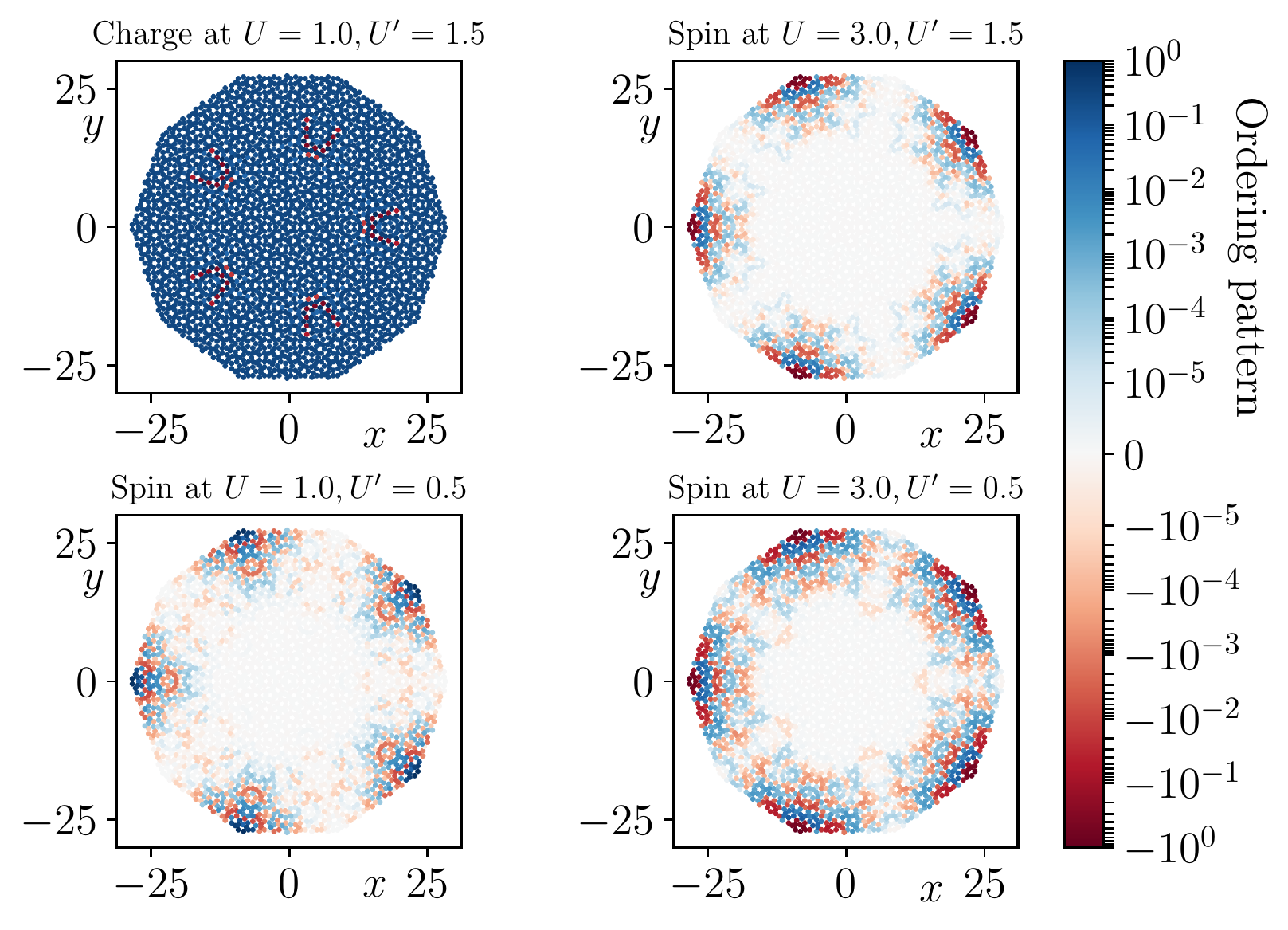}}
\hfill
{\includegraphics[width = 0.32\linewidth]{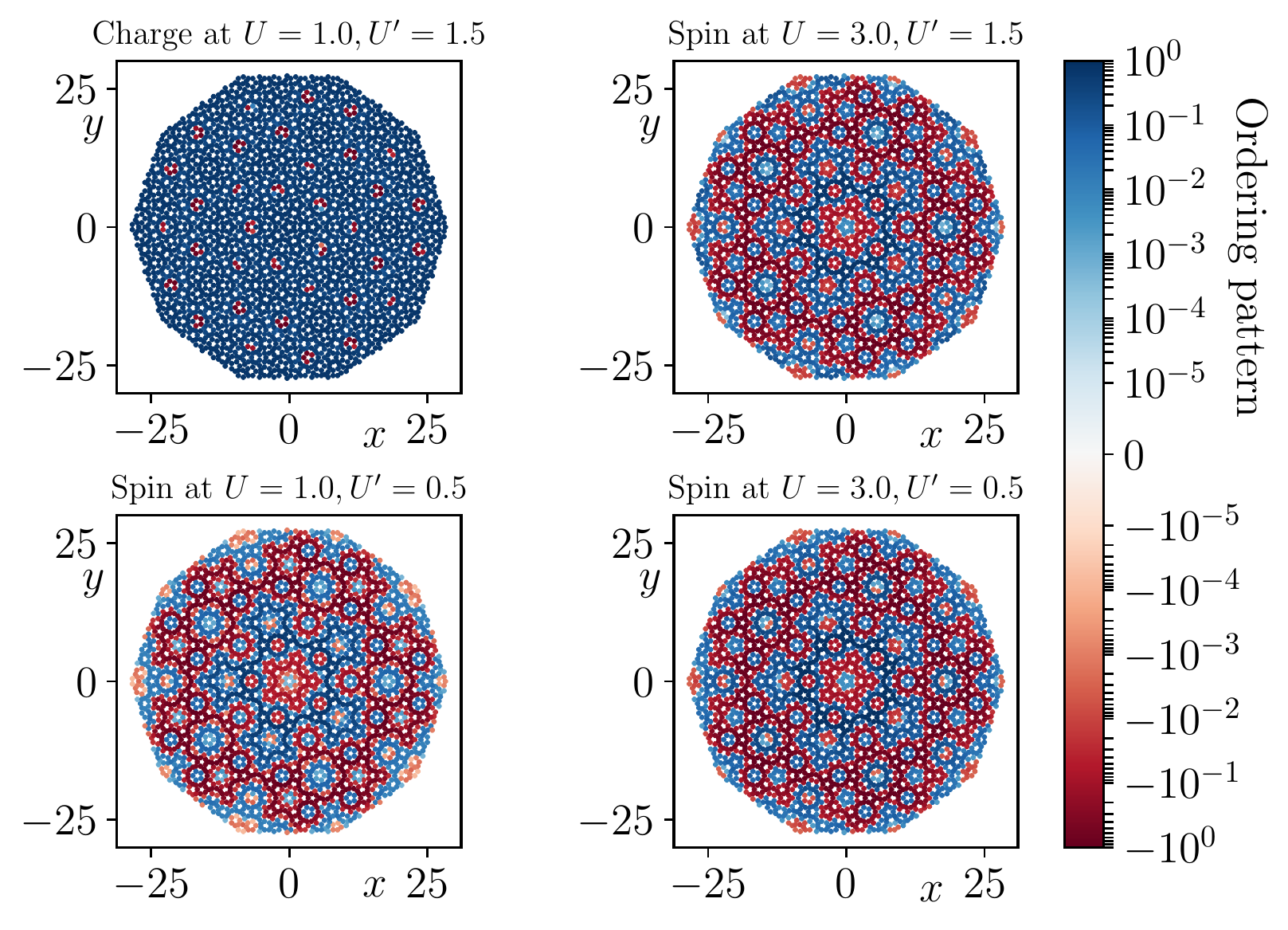}}
\hfill
{\includegraphics[width = 0.32\linewidth]{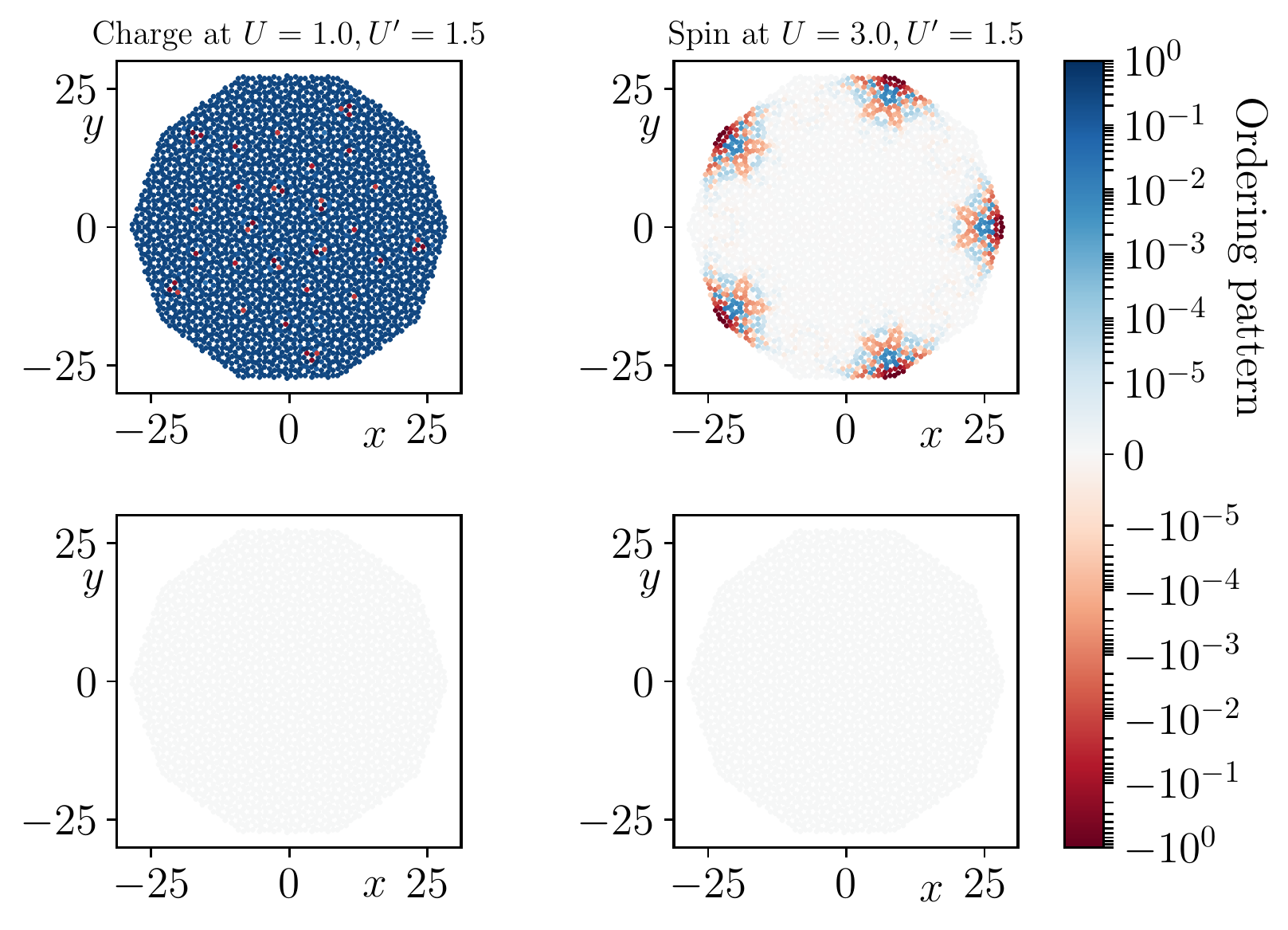}}
\vfill
{\includegraphics[width = 0.32\linewidth]{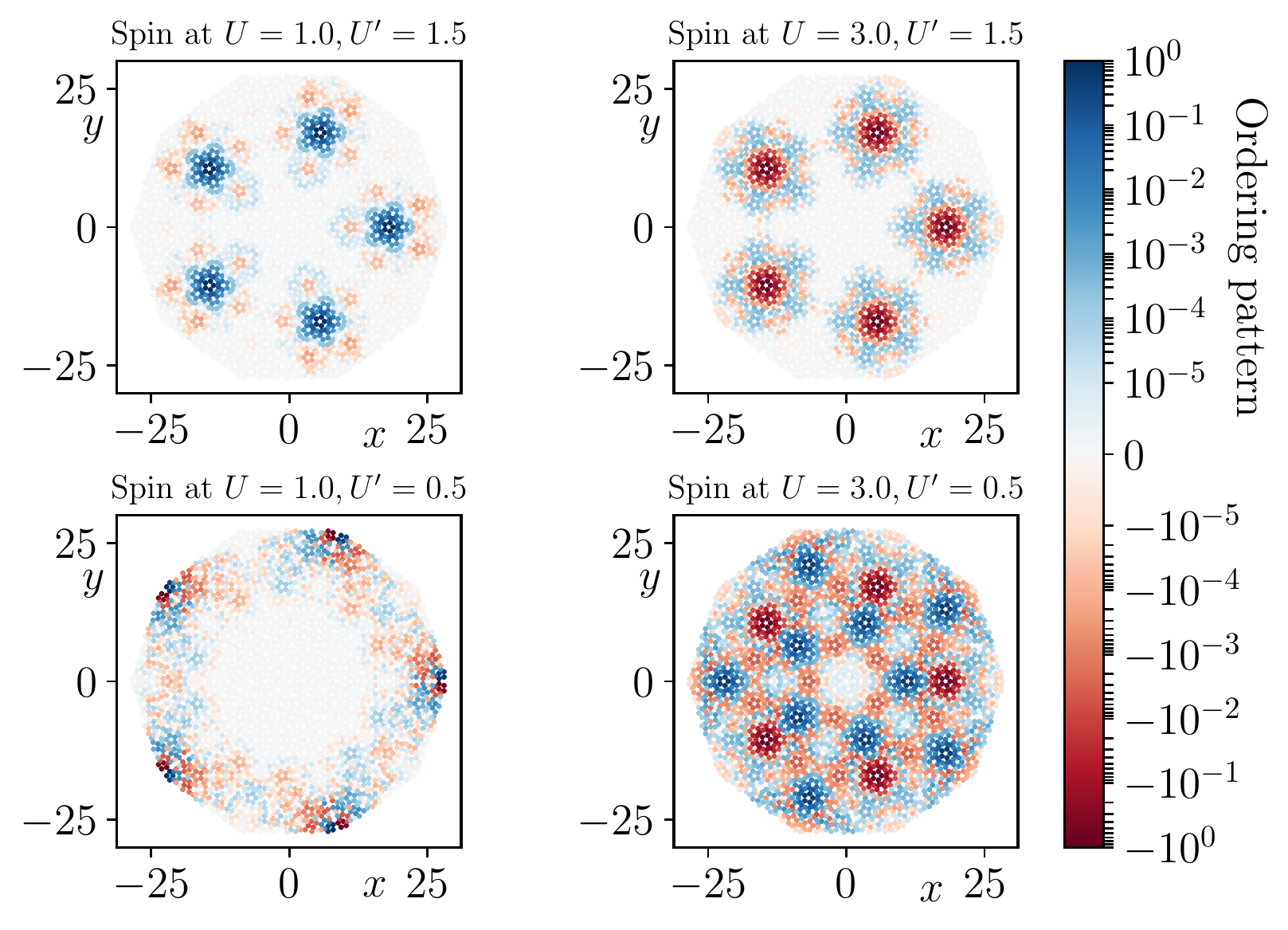}}
\hfill
{\includegraphics[width = 0.32\linewidth]{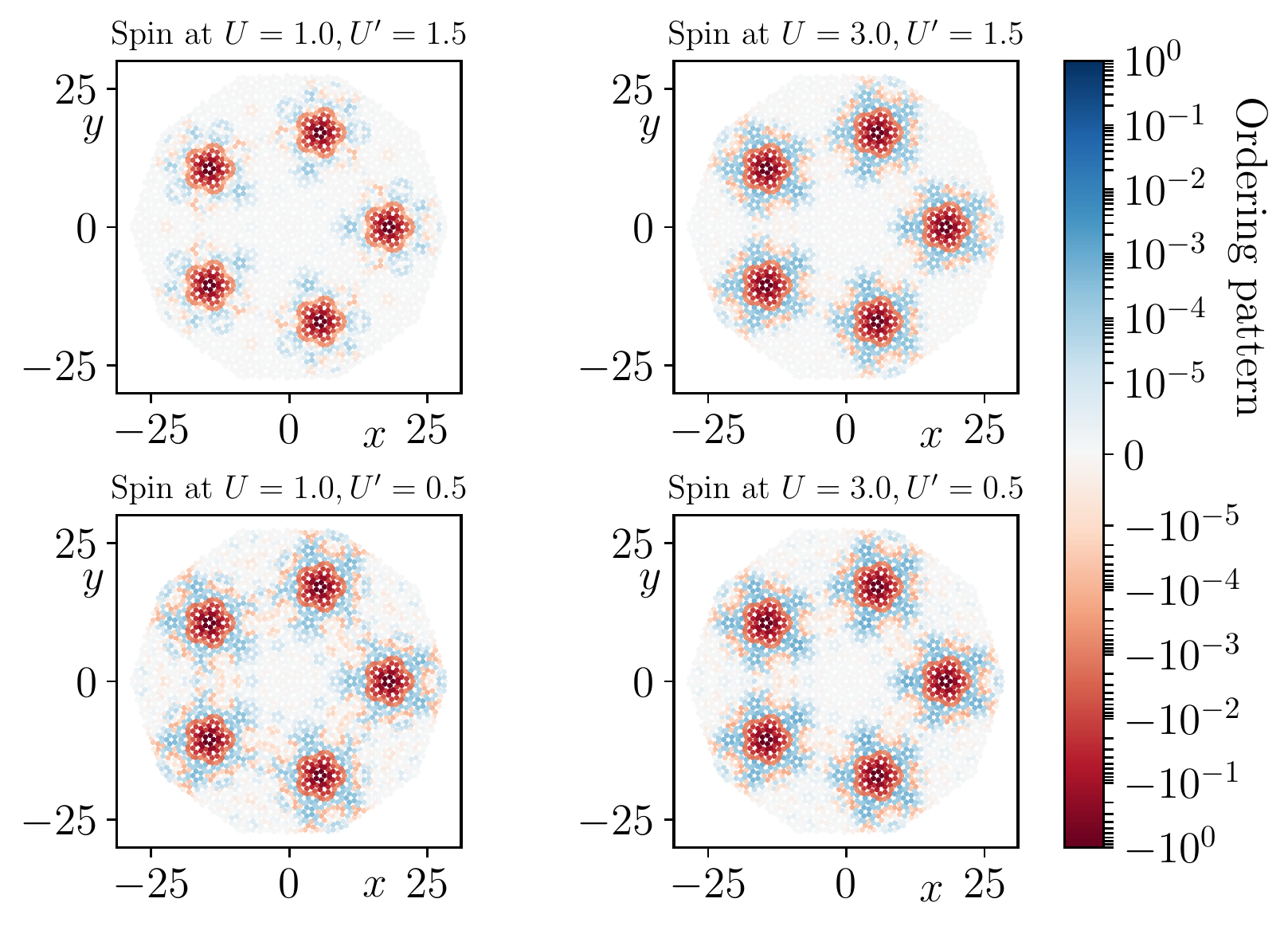}}
\hfill
{\includegraphics[width = 0.32\linewidth]{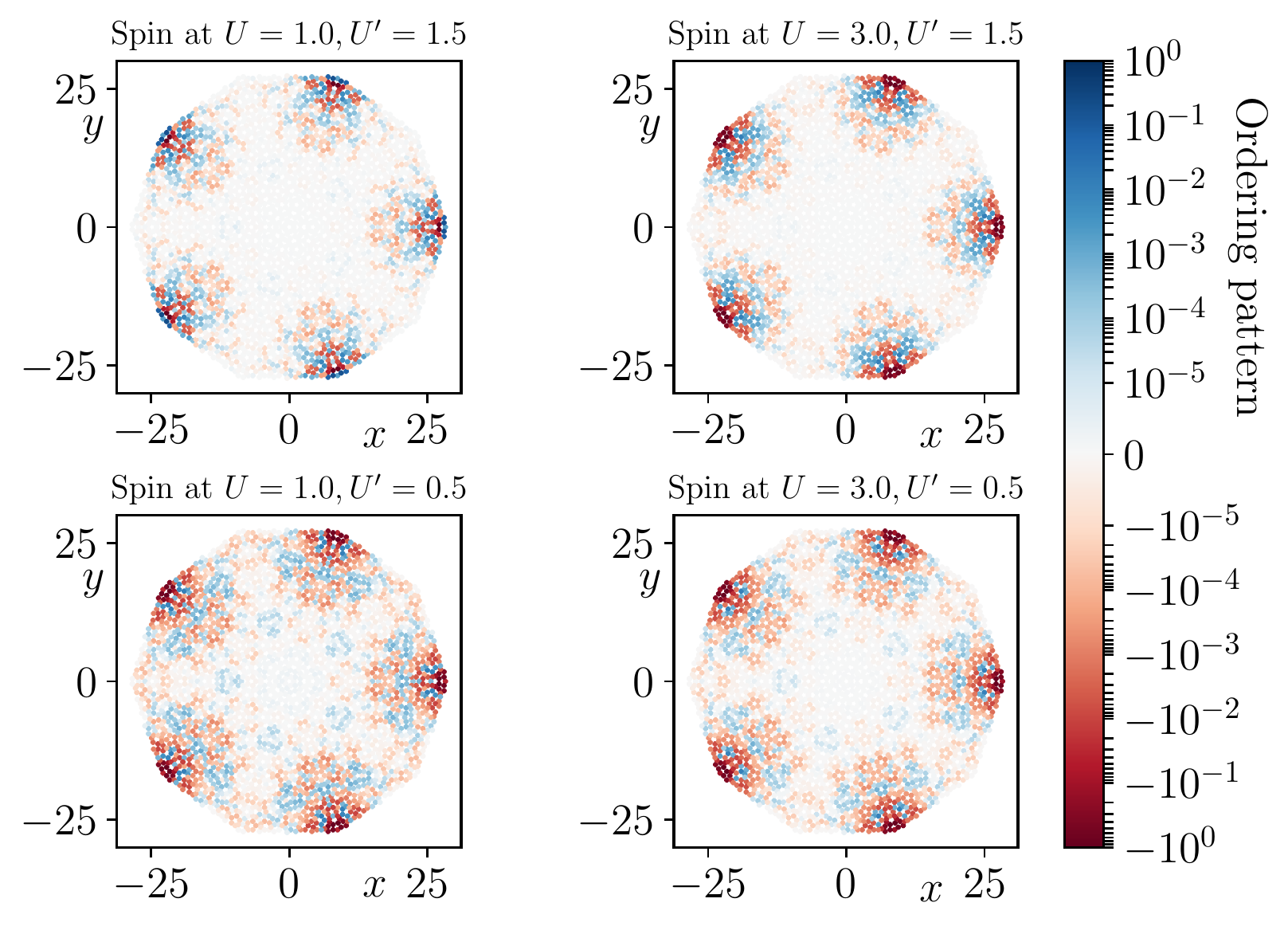}}

\caption{Ordering predictions for each some chosen data-points at each chemical potential, where we show $\mu = 1.9$ in the upper left, $\mu = 2.0$ in the upper middle, $\mu = 2.1$ in the upper right, $\mu = 2.3$ in the lower left, $\mu = 2.35$ in the lower middle and $\mu = 2.4$ in the lower right plot, from FRG plus post-production mean field theory at chosen data points. $U$ and  $U'$ are varied at $T=10^{-3}$, nearest neighbors are included in the calculation.}
\label{Fig::phases_center_mod}
\end{figure}

In Fig.~\ref{Fig::phases_center} and Fig.~\ref{Fig::phases_center_mod} the dependence of the ordering and the phase on $\mu$, $U$ and $U'$ is summarized for the center model. We find that non-localized divergences only occur in the vicinity of the main divergence. For all data sets we found more than a single diverging  eigenvalue, thus more than a single channel mean-field decoupling is performed to extract the ordering information. 

\section{Phase-diagram exponential hopping center model at different filings.}
\label{App:G}
\begin{SCfigure}[][!htbp]
    \centering
    \includegraphics[width = 0.4\textwidth]{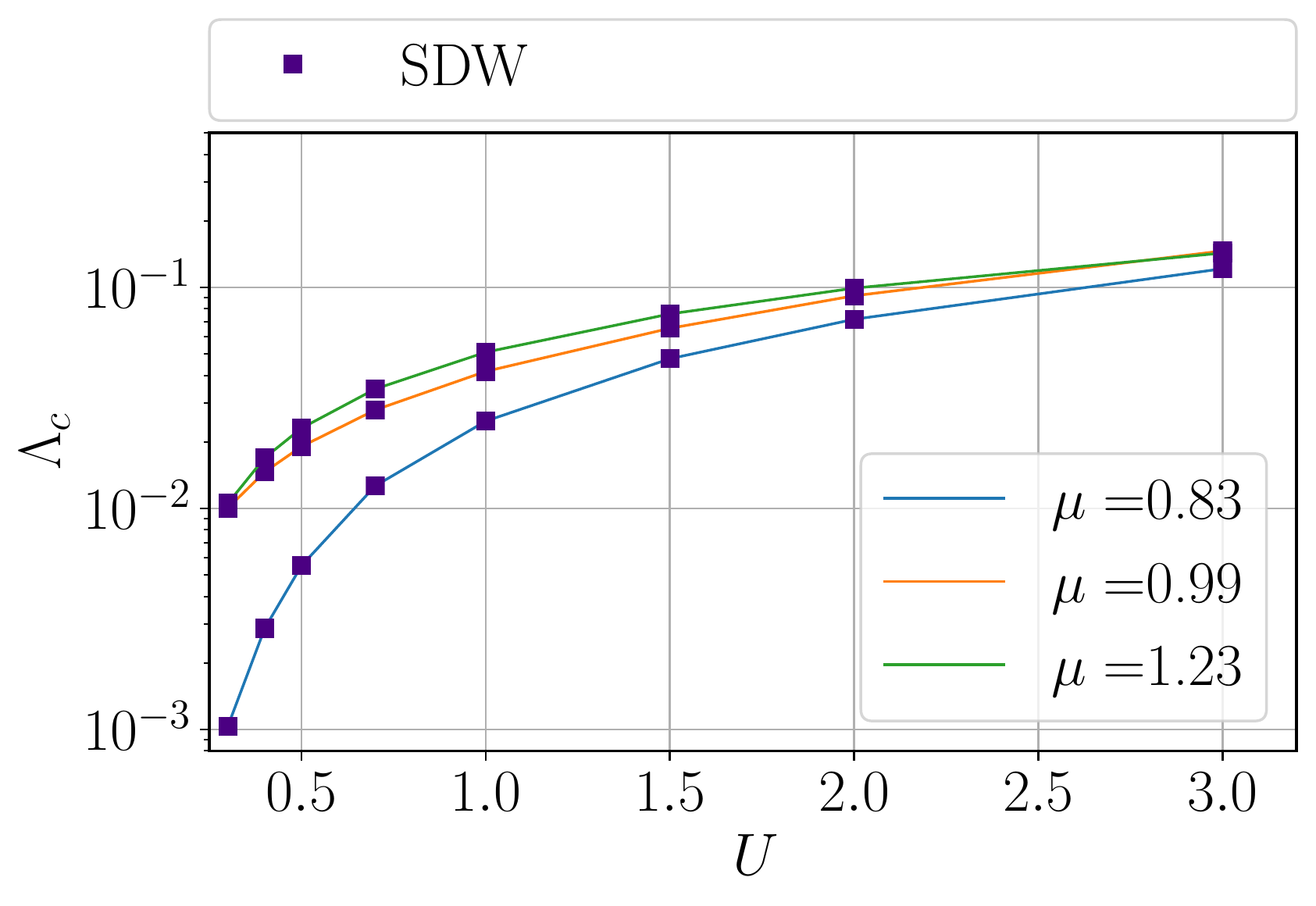}
    \caption{Critical scales depending on the interaction strength in the center model with exponential hoppings at $T=10^{-3}$ including all sites with a distance $\leq3a$ in the calculation.}
    \label{fig:exp_U_dep}
\end{SCfigure}

\begin{figure}[!htbp]
    \centering
    {\includegraphics[width = 0.32\linewidth]{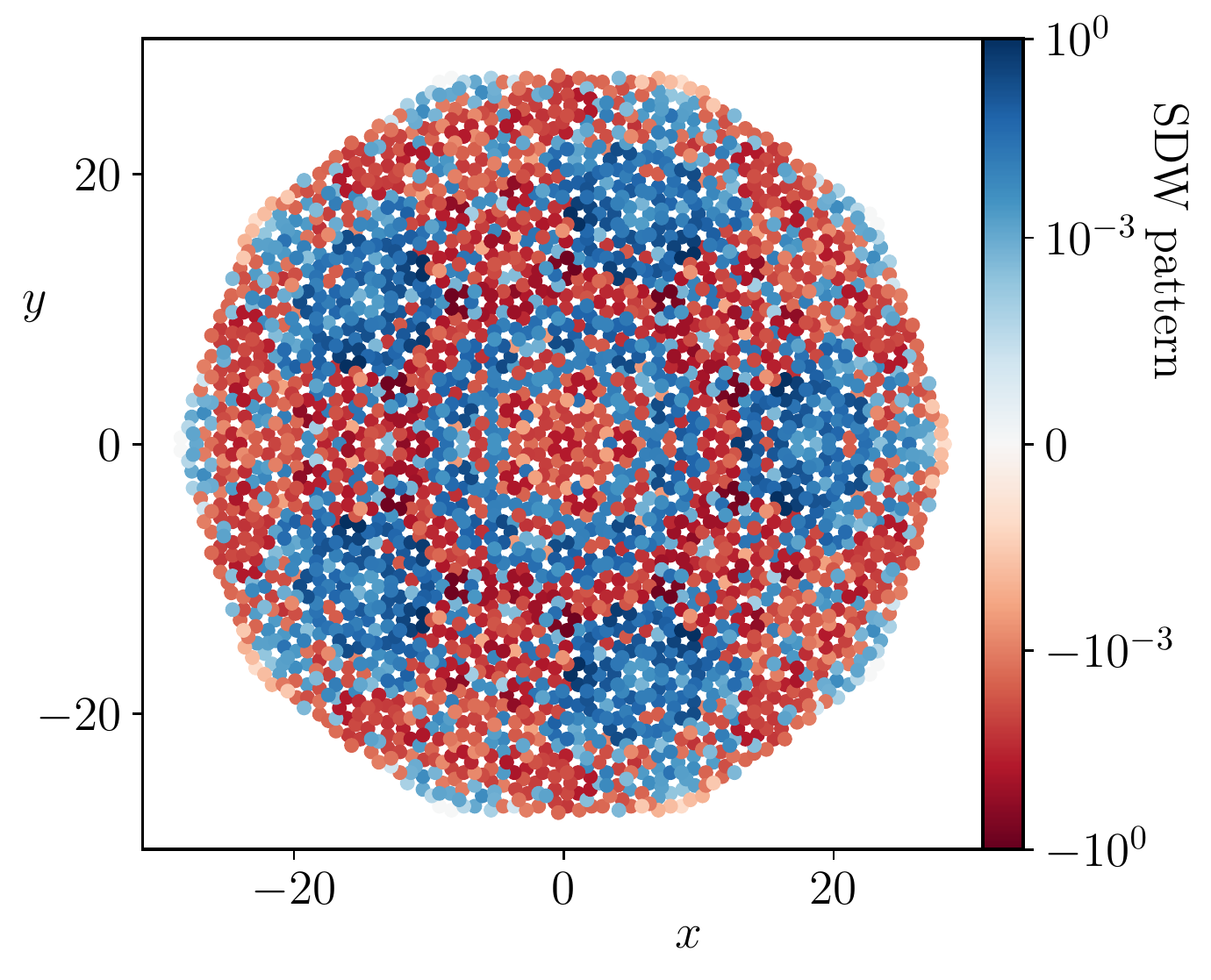}}
    \hspace{1cm}
    {\includegraphics[width = 0.32\linewidth]{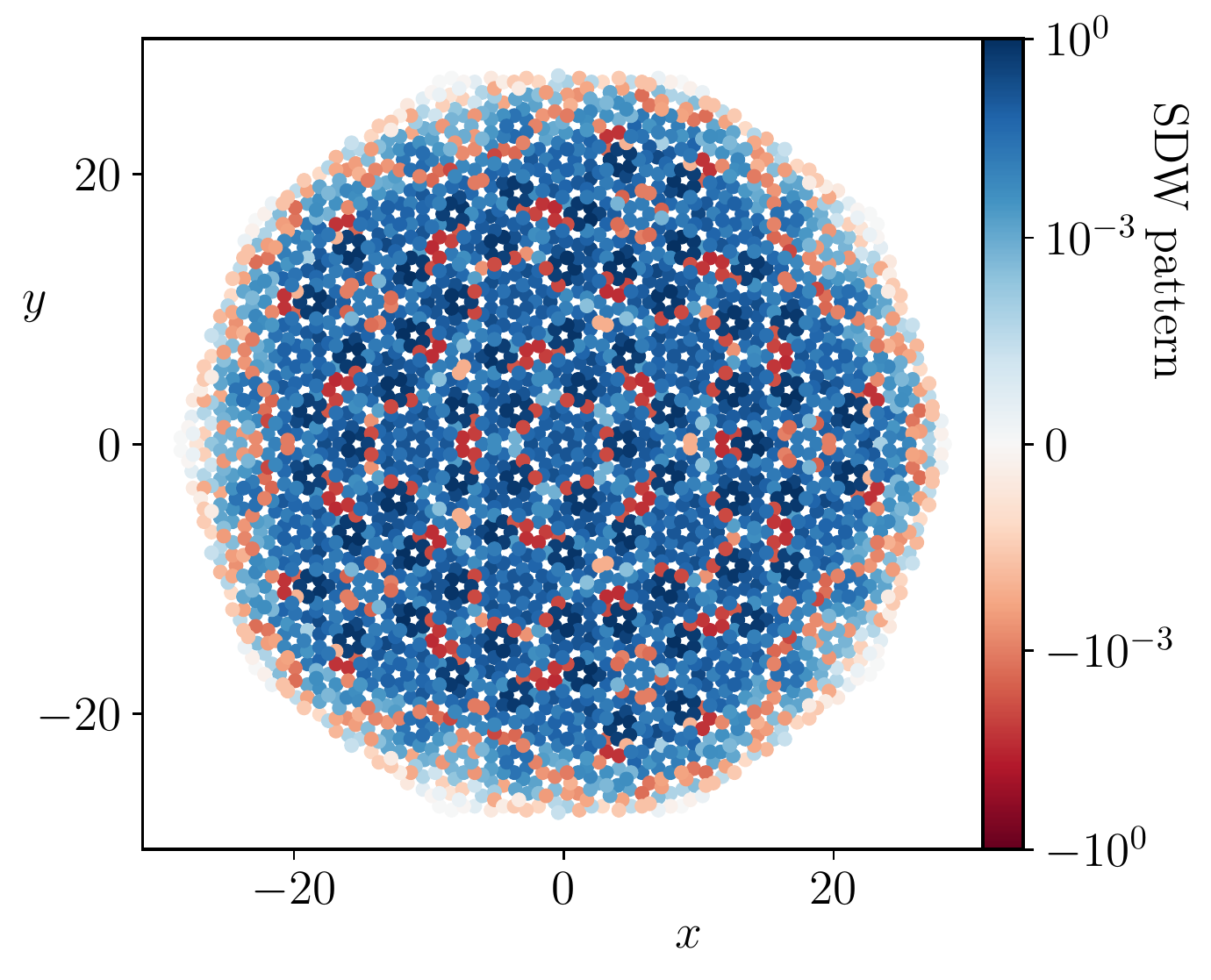}}
    \vfill
    {\includegraphics[width = 0.32\linewidth]{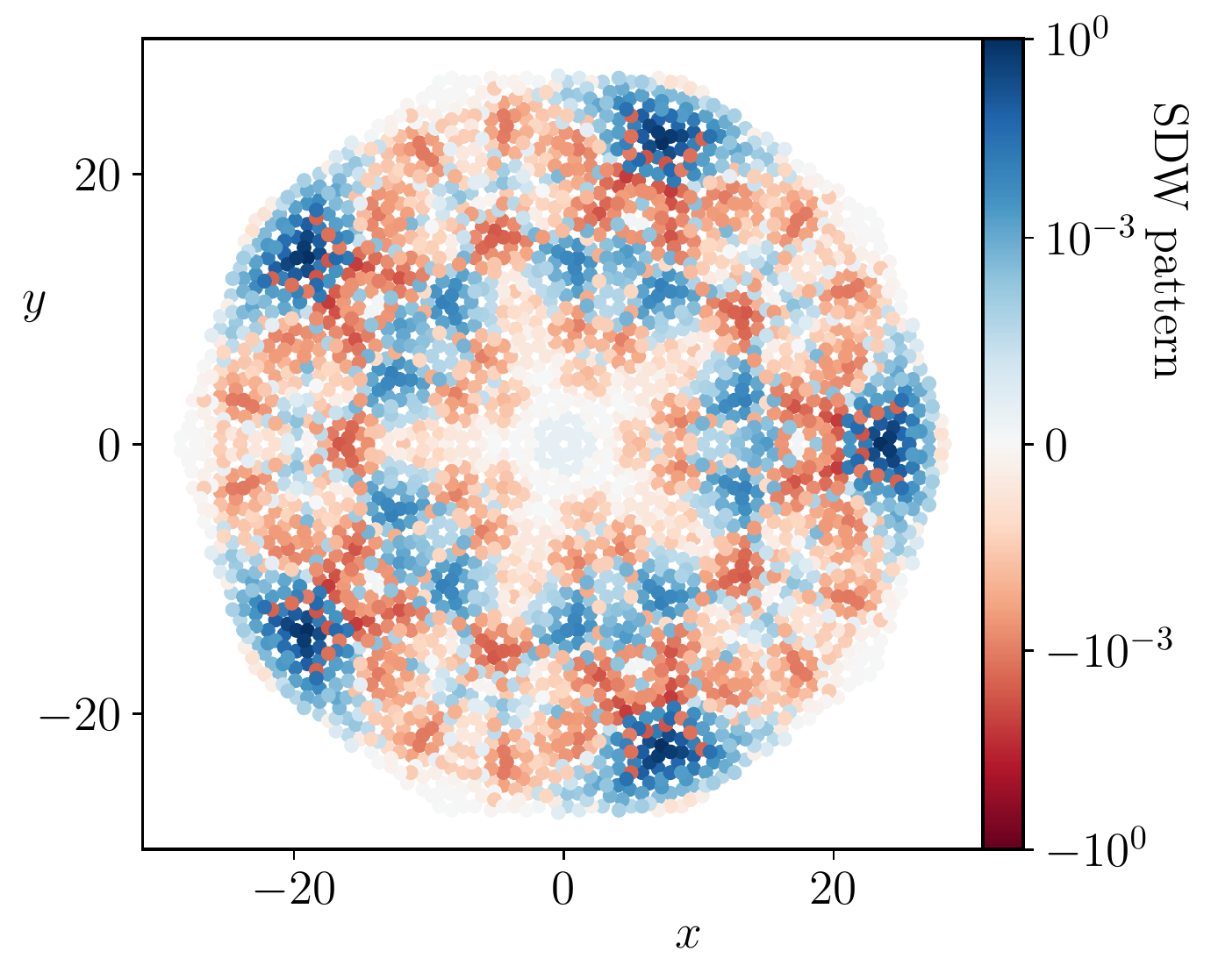}}
    \hspace{1cm}
    {\includegraphics[width = 0.32\linewidth]{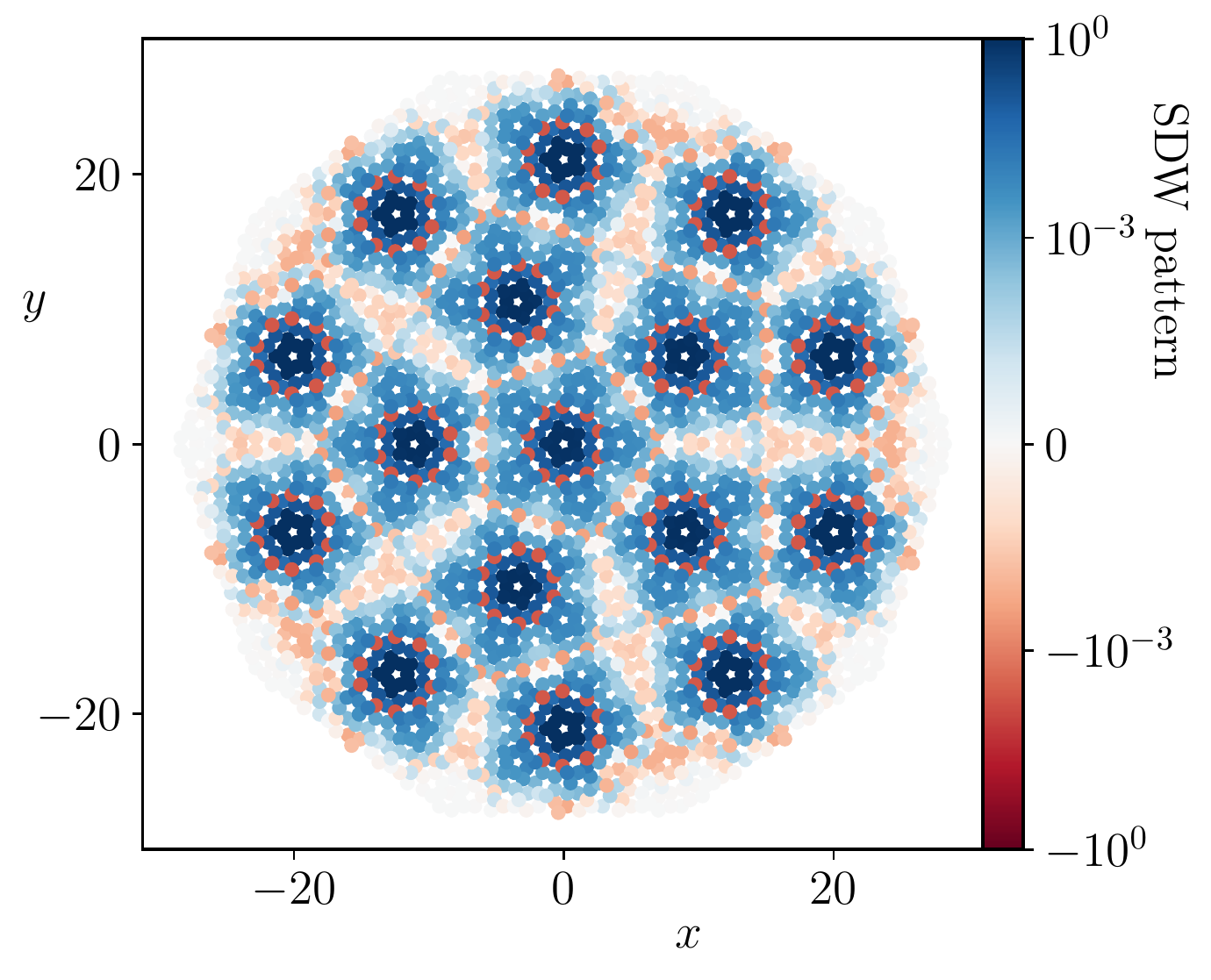}}
    \vfill
    {\includegraphics[width = 0.32\linewidth]{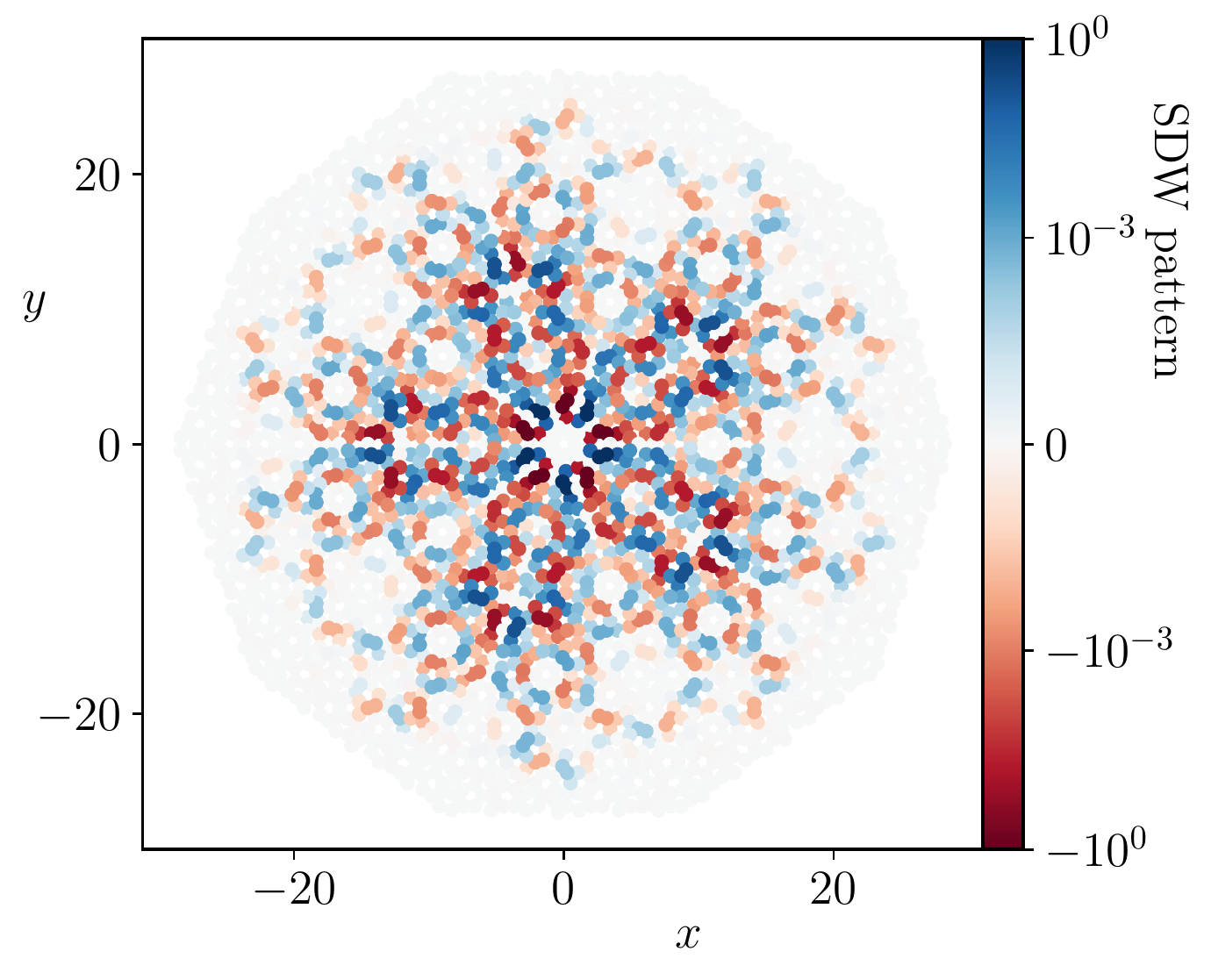}}
    \hspace{1cm}
    {\includegraphics[width = 0.32\linewidth]{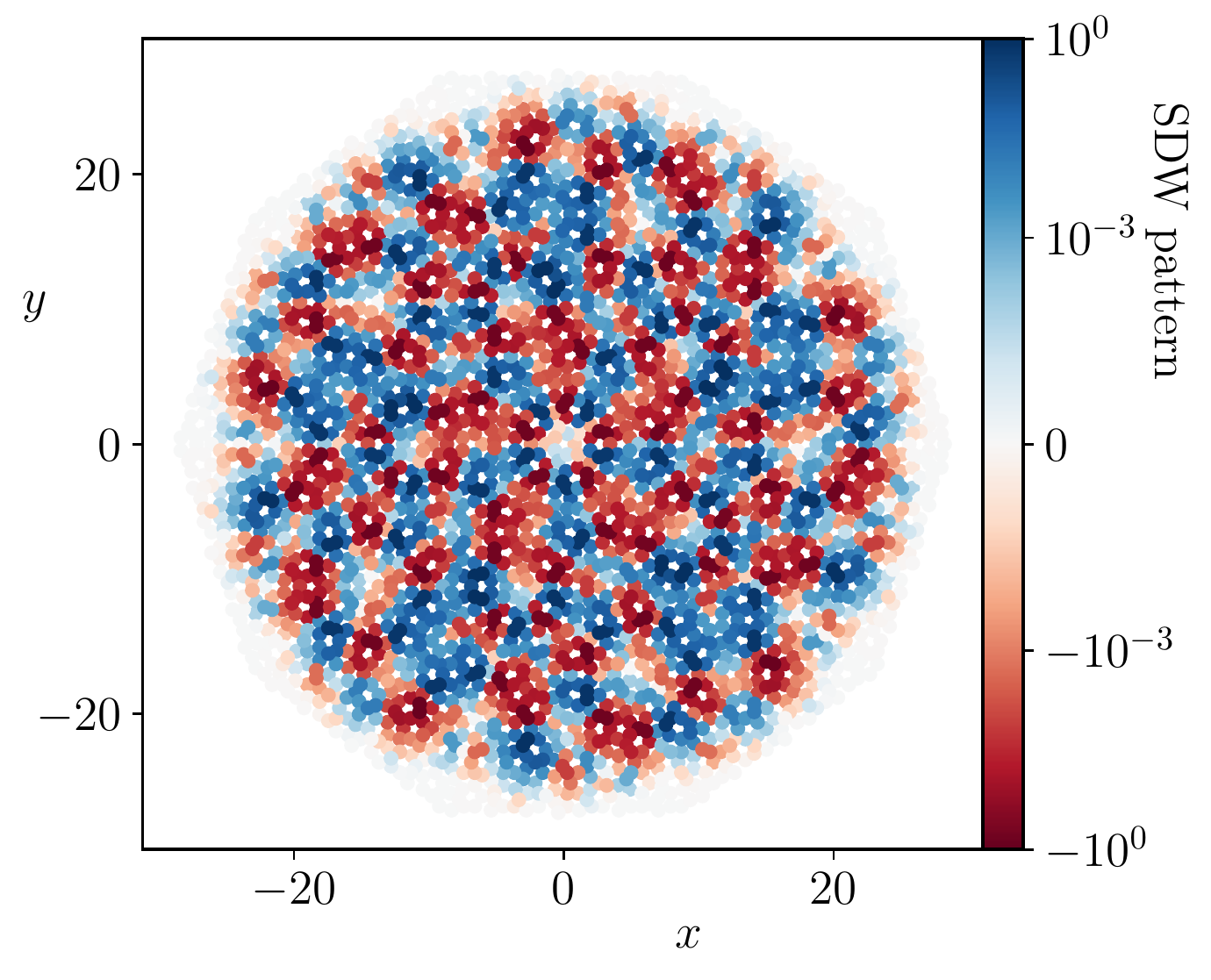}}
    \caption{Orderings at varying chemical potentials and interaction strength, where the left column show orderings for $U = 0.3$ and the right one shows orderings for $U=3.0$ and the chemical potential is shown from lowest ($\mu=0.83$) in the top row, to the highest in the bottom row ($\mu = 1.23$), for the physical center model. Calculations are performed at $T=10^{-3}$ using a frequency-cutoff including all sites with a distance $\leq3a$ in the calculation. The colormap is diverging thus blue and red mark different signs, we normalized the gap to $\pm 1$ and applied a logarithmic scale.}
    \label{fig:my_label}
\end{figure}

For the exponential hopping center model we find the interaction dependence of the critical scale as summarized in Fig.~ \ref{fig:exp_U_dep}. A clear classification in standard categories like AFM and FM is not possible thus we just state that SDW indicates a magnetic ordering.
For each filling we display examples of the SDW ordering at the lowest and highest interaction strength in Fig.~\ref{fig:my_label}. 
\end{document}